\documentclass[twocolumn,showpacs,preprintnumbers,amsmath,amssymb,superscriptaddress,prl]{revtex4-1}

\usepackage{geometry,amsmath,amsfonts,amssymb,array}
\usepackage{graphicx,hyperref,empheq}
\def\p{\partial}
\def\bx{\bgroup \bf x\egroup}
\def\bn{\textbf{n}}
\def\bi{\textbf{e}}
\def\bm{\textbf{m}}

\def\bA{\textbf{A}}
\def\const{\rm const}
\def\tr{\rm tr}

\def\vp{\varphi}

\def\be{\begin{equation}}
\def\ee{\end{equation}}

\let\Im\I

\begin{document}

\title{Forming a cube from a sphere with tetratic order}

\author{O. V. Manyuhina} 
\email{omanyuhi@syr.edu}
\affiliation{Physics Department, Syracuse University, Syracuse, NY 13244, USA}

\author{M. J. Bowick}
\email{bowick@physics.syr.edu}
\affiliation{Physics Department, Syracuse University, Syracuse, NY 13244, USA}
\date{\today}

\begin{abstract}
Composed of square particles, the tetratic phase  is characterised by a four-fold symmetry with quasi-long-range orientational order but no translational order. We construct the elastic free energy for tetratics and find a closed form solution for $\pm1/4$ disclinations in planar geometry. Applying the same covariant formalism to a sphere we show analytically that within the one elastic constant approximation eight $+1/4$-disclinations favor positions defining the vertices of a cube. The interplay between defect--defect interactions and bending energy results in a flattening of the sphere towards superspheroids with the symmetry of a cube.
\end{abstract}

\maketitle

%

Nature provides many fascinating examples of self-assembly at a variety of length scales leading to rather complex morphologies~\cite{book:growth}.  Yet we are still far from understanding the basic design principles governing self-assembly and from predicting the resulting equilibrium shapes based on both microscopic structure and the detailed form of the interactions. Establishing the connection between different length scales is not only of fundamental interest but also important for engineering new (nano)materials in a controlled way with {\it a priori} known shape and mechanical properties. An insightful approach to study the interplay between order and geometry involves topological defects, that, on the one hand, capture discrete symmetries of the constituent building blocks and, on the other hand, are relevant degrees of freedom to account for global topological constraints and  boundary conditions. The interplay between the elastic anisotropy of liquid crystals and the bending rigidity of a sphere, for example, results in the variety of morphologies predicted in~\cite{bowick:pnas}, most strikingly the faceted tetrahedron,  and observed experimentally in vesicles self-assembled from block copolymers. Indeed, within the one elastic constant approximation the ground state of a nematic liquid crystal confined to a spherical surface has four $q_i=+1/2$ disclinations placed at the vertices of tetrahedron inscribed in the sphere~\cite{lubensky:1992,park:1992,vitelli:2006,bowick:review}. Because of the long-range repulsive nature of interactions, topological defects maximize their geodesic distance on a sphere. The authors of~\cite{lubensky:1992,park:1992} conjectured that the ground state for the tetratic phase with four-fold rotational symmetry confined to a closed surface includes eight  $+1/4$ disclinations at the vertices of a twisted cube (opposite faces of a cube are rotated by $\pi/4$). Though recent Monte-Carlo simulations~\cite{chen:2013} confirmed this long-standing hypothesis the precise form of the free energy for the tetratic phase and the resultant ground states remains open~\cite{chaikin:2007,selinger:2009,donev:2006}. 


In this Letter we propose a phenomenological description of tetratic order and construct the simplest form of the free energy invariant under the operations of the dihedral point symmetry group $D_{4h}$~\cite{book:group}. We consider the limiting cases of this model and describe tetratic configurations in the plane and on the two-sphere. It turns out that eight $+1/4$ disclinations energetically prefer the vertices of cube rather than the twisted cube with maximum separation between defects. Hence, in presence of tetratic order, a sphere with finite bending rigidity may deform towards a rounded cube, which is a promising building block for self-assembly of larger structures~\cite{glotzer:2007} or perhaps for drug delivery to a tumour~\cite{black:2014}.

\begin{figure}[tb]
\centering
\includegraphics[width=0.95\linewidth]{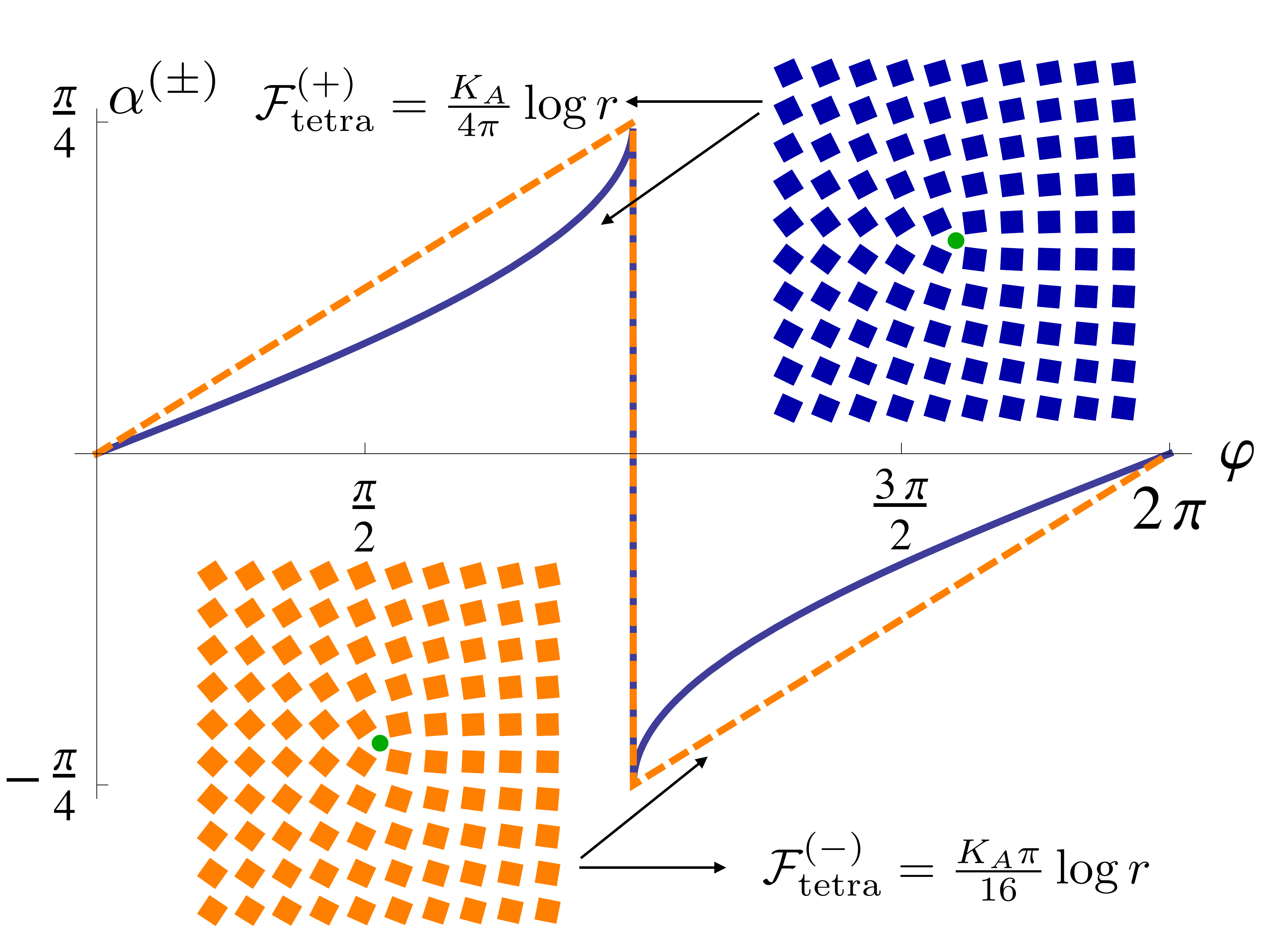}
\caption{\label{fig:defect14}(Color online) The change of the angle $\alpha^{(+)}$~\eqref{eq:arcsin} upon encircling a $+1/4$ disclination (solid line, top configuration) and of the angle $\alpha^{(-)}=1/4\Im\log(re^{i\vp})$ (dashed line, bottom configuration). The tetratic model$^{(+)}$ gives the energetically favourable solution with ${\cal F}_{\rm tetra}^{(+)}/{\cal F}_{\rm tetra}^{(-)}=4/\pi^2$~\eqref{eq:f14}.}
\end{figure}

The square is the elementary constituent of the tetratic phase.  In planar geometry, it can be characterised by two unit vectors $\bn\leftrightarrow-\bn$ and $\bm\leftrightarrow -\bm$, which are orthogonal ($\bn\cdot\bm=0$) and transform into each other under $\pi/2$ rotations. The relevant order parameter for the tetratic phase is a fourth-rank symmetric traceless tensor $\propto n_in_jn_kn_l$, where $i,j,k,l=1,2$~\cite{park:1992, see:SM}. Following the methods used in constructing the Frank free energy for nematic liquid crystals~\cite{frank:1958}, we construct  the elastic free energy for the tetratic phase, by assuming that it is quadratic in gradients of the director $\bn$ up to the lowest order $f_{\rm tetra}\propto \lambda_{ijklsp}\,n_i n_j \nabla_k n_l \nabla_s n_p$. We find only four independent symmetry allowed terms in $f_{\rm tetra}$~\cite{see:SM}, invariant under the operations of the dihedral point symmetry group ${\cal D}_{4h}$~\cite{book:group}. Here we consider two limiting cases, assuming either the one constant approximation or the vanishing of $\lambda_{121112}=\lambda_{122221}=0$, yielding $f_{\rm tetra}^{(\pm)}\propto\sum_{\substack {i=1\\i\neq j}}^2\big (n_i (\nabla\bn)_{ji}\pm n_j(\nabla\bn)_{ii}\big)^2$ (see Eqs.~(16$^*$), (17$^*$) in SM~\cite{see:SM}). Parametrizing the director in a local orthonormal system of coordinates as $\bn=\cos\alpha\,\bi_1+\sin\alpha\,\bi_2$ we write the tetratic  free energy density in the form
\be\label{eq:fpm}
f_{\rm tetra}^{(\pm)}=\frac {K_A}2\left\{ \begin{aligned} \cos^2(2\alpha)|\nabla\bn|^2,\quad &\mbox{(one constant)},\\ |\nabla\bn|^2,\quad &\mbox{(nematic-like)}. \end{aligned}\right.
\ee
Reducing the number of phenomenological elastic constants $\lambda$ allows us to explore the proposed models analytically. Note that the model$^{(+)}$ is more generic and accounts for the fourfold term proportional to $\cos(4\alpha)$. Introducing such term into the lattice model~\cite{selinger:2009} results into the phase transition from a nematic or isotropic phase to a tetratic phase.

The behavior of vortices ($|q_i|=1$) in superfluids and defects in nematic liquid crystals ($|q_i|=1/2$) is conventionally studied within the $XY$model~\cite{vitelli:review,nussinov:2014} and Frank free energy~\cite{bowick:review,vitelli:2006}, respectively. They are analogous to $f_{\rm tetra}^{(-)}$, whose solution of the Laplace equation $\Delta \alpha^{(-)}=0$ may account for an arbitrary jump of the angle $\alpha$ by $2\pi q_i$ across the branch cut. For tetratics the angle $\alpha^{(-)}(z)=1/4 \Im\log(z)$, $z=x+iy=re^{i\vp}$ changes by $\pi/2$ as we go around the disclination $\vp\in(0,2\pi)$ at $r=0$, giving the charge  $q_i=+1/4$. The solution of the Euler--Lagrange equation associated with~$f_{\rm tetra}^{(+)}$ is~\cite{see:SM}
\be\label{eq:arcsin}
\alpha^{(+)}(z)=\frac 12 \arcsin\bigg\{\frac {\Im\log(z)}\pi\bigg\}, 
\ee
accounts as well for an isolated disclination of charge $q_i=+1/4$. In Fig.~\ref{fig:defect14} we compare these two solutions and list the  free energies associated with $+1/4$ disclination 
\be\label{eq:f14}
{\cal F}_{\rm tetra}^{(+)}=\frac{K_A}{4\pi}\log(r),\quad  {\cal F}_{\rm tetra}^{(-)}=\frac{K_A \pi}{16}\log(r),
\ee
found by integrating~\eqref{eq:fpm}. Interestingly, the model$^{(+)}$ (one constant) has $4/{\pi^2}$ times lower free energy than the model$^{(-)}$ (nematic-like). Thus, we believe that the proposed description of the tetratic phase is relevant to the study of equilibrium structures, as discussed below.

\begin{figure}[t]
\centering
\raisebox{33mm}{(a)}\includegraphics[width=0.43\linewidth]{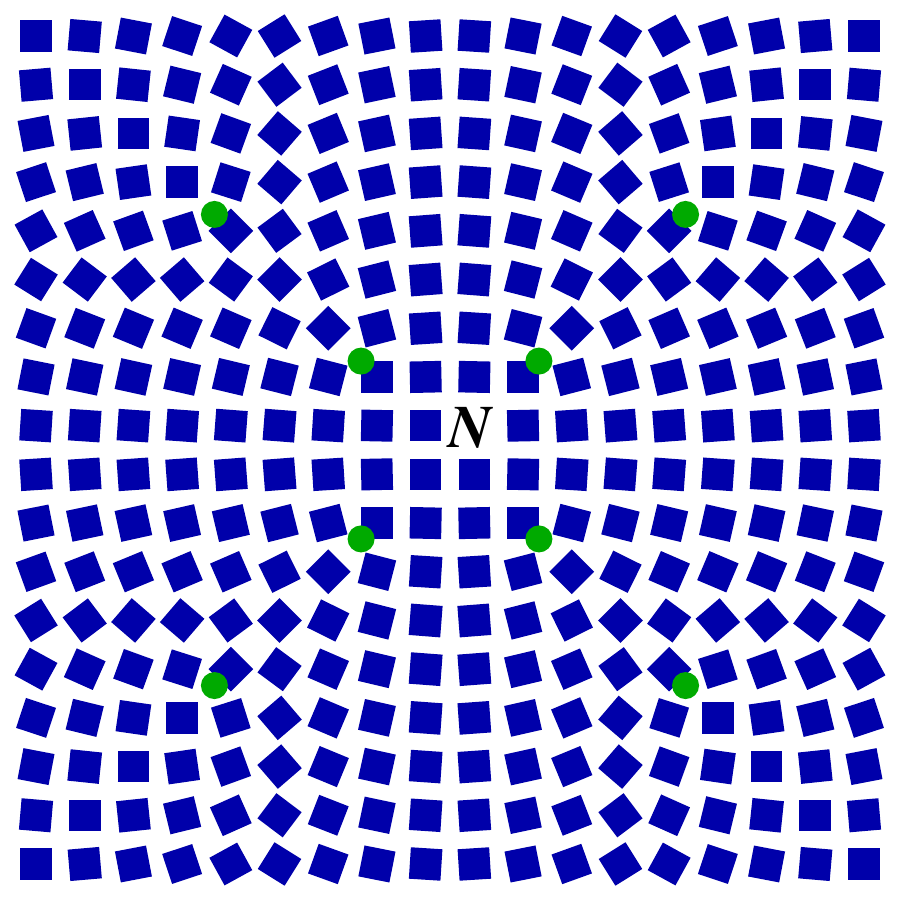}
\hfil
\raisebox{33mm}{(b)}\includegraphics[width=0.43\linewidth]{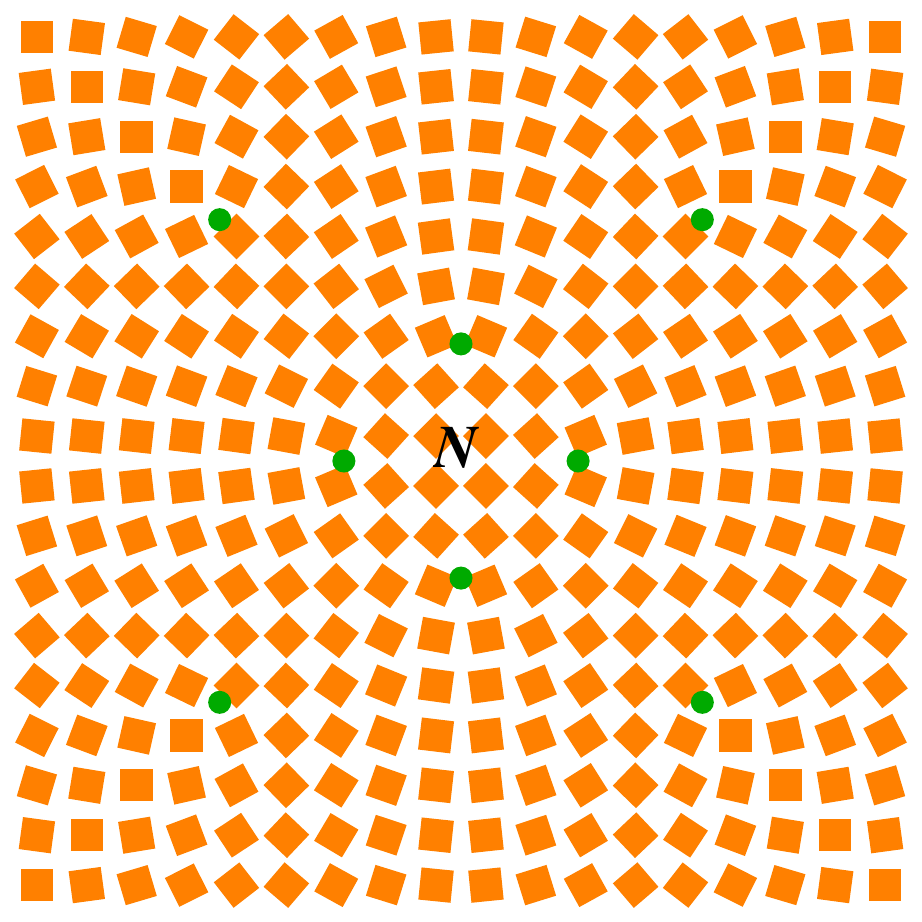}
\vskip0.1ex
\raisebox{33mm}{(c)\kern-10pt}\includegraphics[width=0.43\linewidth]{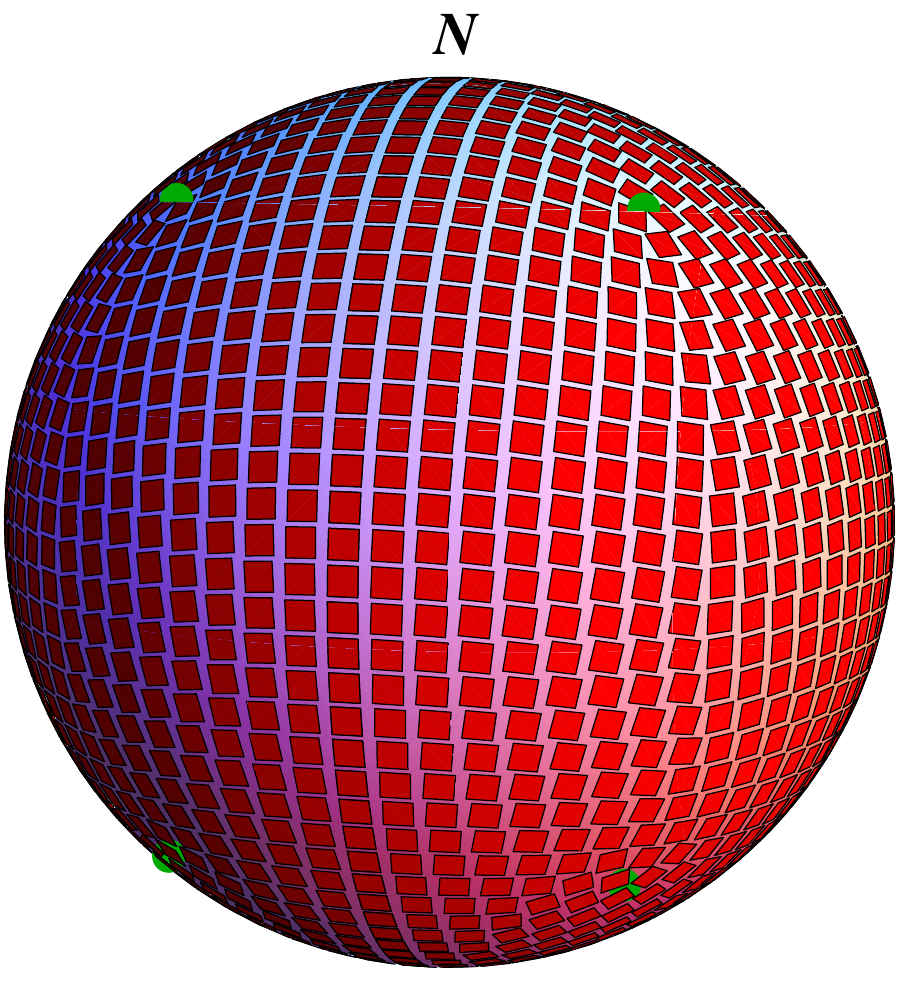}
\hfil
\raisebox{33mm}{(d)\kern-10pt}\includegraphics[width=0.43\linewidth]{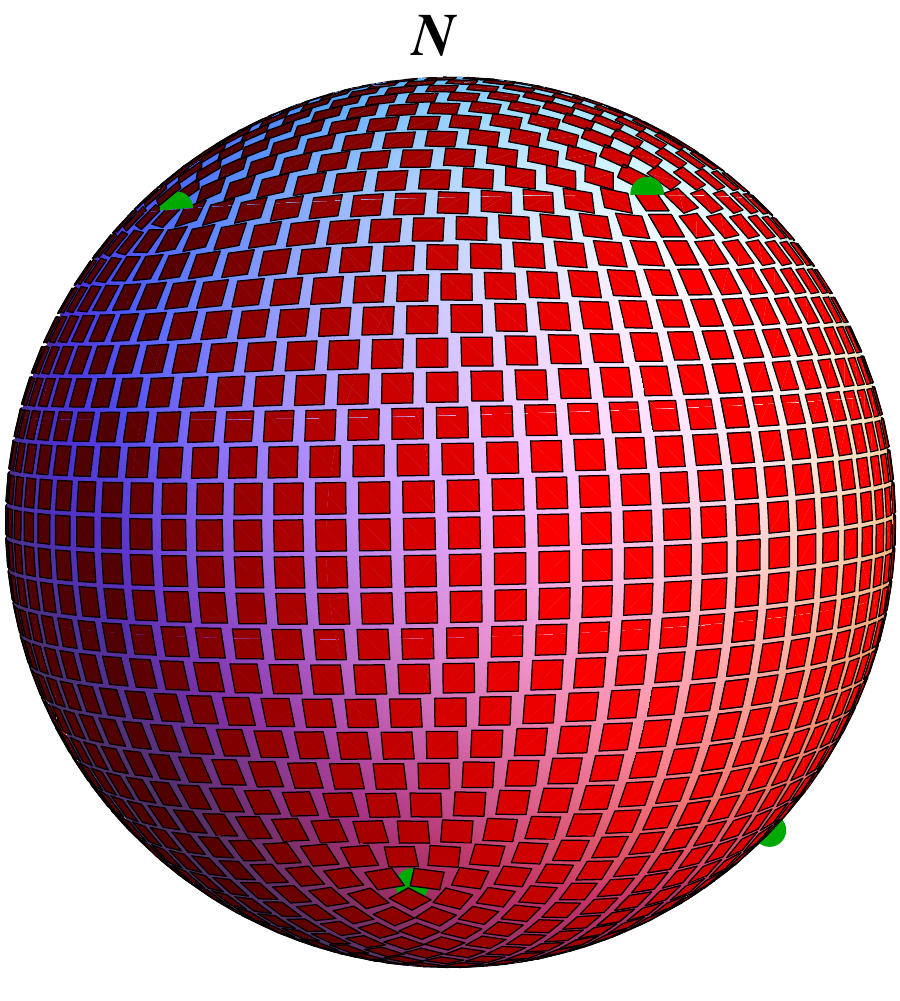}
\vskip1ex
\raisebox{27mm}{(e)\kern10pt}\includegraphics[width=0.65\linewidth]{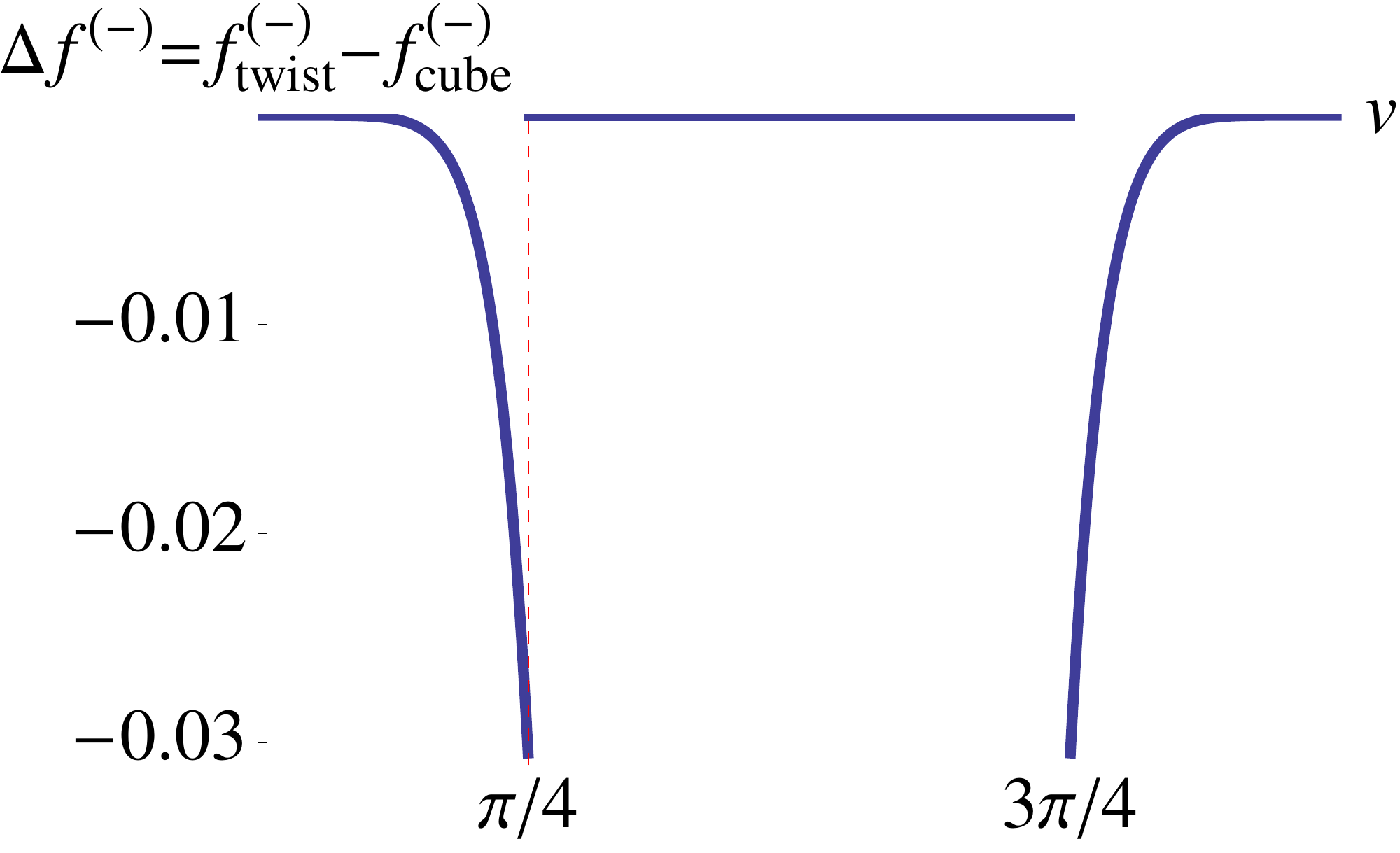}
\caption{\label{fig:sphere}(Color online) Tetratic order: on a projected plane $z=x+iy$ with 8 (+1/4) disclinations~\eqref{eq:proj} at the vertices of two concentric squares (a) and twisted by $\pi/4$ (b); stereographic projection on a sphere  (a)$\to$(c) and (b)$\to$(d), similar to~\cite{lubensky:1992}. (e) The free energy difference between (d) and (c) configurations within the model$^{(-)}$ (nematic-like), the integrand $\pm 8\pi \tan^7(\frac v2)/\big((1+\cos v)(1-\tan^{16}(\frac v2))\big)$~\cite{see:SM}.}
\end{figure}

Consider tetratic order on a sphere. We expect eight disclinations of $q_i=+1/4$ charge giving total charge $\sum q_i=+2$, the Euler--Poincar\'e characteristic of a two-sphere~\cite{docarmo}. An earlier theoretical prediction in~\cite{lubensky:1992} and recent Monte Carlo simulations~\cite{chen:2013} suggest that the energy minimizing configuration of eigth disclinations correspond to the vertices of a twisted cube inscribed into a sphere. In the following we analyze this hypothesis and quantify the energetics of defects within the proposed models$^{(\pm)}$.

The tetratic free energy~\eqref{eq:fpm} on a two-dimensional surface ${\cal S}$ in ${\mathbb R}^3$, endowed with orthogonal coordinates $(u,v)$ and the metric $ds^2=E \,du^2+G\,dv^2 $, can be written by replacing the gradient of $\bn$ by~\cite{napoli:2012,docarmo,see:SM}
\be\label{eq:neg}
|\nabla\bn|^2= \bigg(\frac{\p_u \alpha}{\sqrt E} -\frac{\p_v E}{2E\sqrt G}\bigg)^2+\bigg(\frac{\p_v\alpha}{\sqrt G} +\frac{\p_uG}{2G\sqrt E}\bigg)^2,
\ee
accounting for the covariant derivatives and geodesic curvatures of ${\cal S}$. For a sphere of radius $R$ we have $E=R^2\sin^2 v$ and $G=R^2$.

\begin{table}[bt]
\caption{\label{tab:ene}Integrated free energy $f_{\rm tetra}^{(\pm)}$~\eqref{eq:fpm} using \eqref{eq:neg} on a sphere for two configurations (Fig.~\ref{fig:sphere}). We choose sphere radius $R=1$ and the core size of defects to be $\delta=0.01$.}
\smallskip
{\def\tabcolsep{10pt}
\def\arraystretch{1.25}
\begin{tabular}{c|c cc}
${\cal F}, K_A$& {\bf geodesics} & {\bf model$^{(-)}$} & {\bf model$^{(+)}$}\\
\hline
{\bf twisted} & 3.7156  & 5.1962 &  4.0535\\[1ex]
{\bf cube} & 3.7183  & 5.1994 &  3.1932
\end{tabular}}	
\end{table}

We compare two configurations with defects located at the vertices of a cube (Fig.~\ref{fig:sphere}a,c) and at the vertices of twisted cube (Fig.~\ref{fig:sphere}b,d). By projecting the sphere stereographically on a complex plane with $z=re^{i\vp}$, so that the meridians ($u=\const$) transform into radial rays ($\vp=\const$) emanating from the North pole ($N$), while the parallels ($v=\const$)  become concentric circles ($r=R\tan(\frac v2)$), we can describe the tetratic order on a complex plane with eight $+1/4$ disclinations  as~\cite{see:SM}
\be\label{eq:proj}
\alpha_{a,b}^{(+)}(z)=\frac 12\arcsin\bigg\{\frac {\Im \log[(z^4\pm r_1^4)(z^4+r_2^4)]} \pi\bigg\},
\ee
where the plus(minus) signs corresponds to Fig.~\ref{fig:sphere}a(b), respectively, and $r_1= R\tan(\frac \pi 8)$, $r_2=R\tan(\frac{3\pi}8)$. Projecting back on a sphere (see Fig.~\ref{fig:sphere}c,d) we get the angle $\alpha_{c,d}^{(+)}(u,v)=\pi/2-\alpha_{a,b}(z)+\arg(z)$ with $z=R\tan(\frac v 2)e^{iu}$. Next we substitute this {\it ansatz}  in~\eqref{eq:neg} and integrate the free energy ${\cal F}_{\rm tetra}^{(+)}=R^2\int\!du\,dv\,\sin v f_{\rm tetra}^{(+)}$ numerically using {\it Mathematica} with the cut-off angle around defects being $\delta/R=0.01$ (last column in Table~\ref{tab:ene}). Within the tetratic model$^{(+)}$ we find that $+1/4$ defects energetically prefer to sit at the vertices defining an inscribed cube, so that the ground state is Fig.~\ref{fig:sphere}c. 

On the other hand, within the model$^{(-)}$ we recover the ground state predicted in~\cite{lubensky:1992,chen:2013}, namely the twisted cube Fig.~\ref{fig:sphere}d. Moreover, we find an analytic expression for the free energy difference between the twisted and straight cube configurations with $\alpha_{c,d}^{(-)}(z)=\pi/2-\frac 14\arg[(z^4\pm r_1^4)(z^4+r_2^4)]+\arg(z)$, plotted in  Fig.~\ref{fig:sphere}e. Integrating over $v$ gives $\Delta{\cal F}_{\rm tetra}^{(-)}/K_A=-\frac \pi4\log(\frac{289}{288})\simeq0.0027$, which coincides with the free energy computed through geodesic distance~\cite{bowick:review,see:SM} (first column in Table~\ref{tab:ene}). The direct numerical integration for model$^{(-)}$ (second column in Table~\ref{tab:ene}) also gives the free energy difference to be less than $1\,\%$ in units of $K_A$.

\begin{figure}[tb]
\centering
\raisebox{45mm}{(a)\kern-5pt}\includegraphics[width=0.65\linewidth]{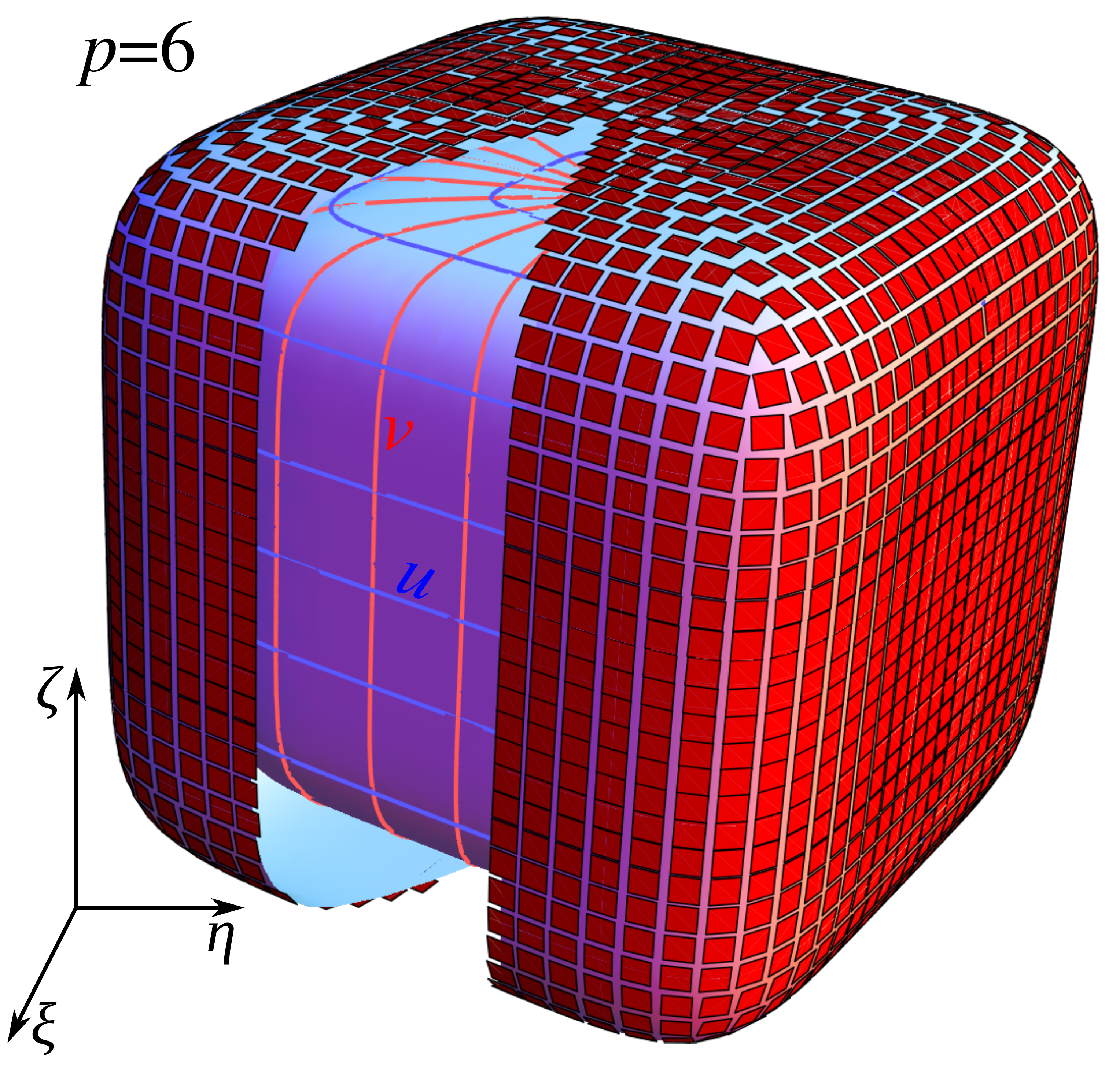}
\vskip0.1ex
\raisebox{45mm}{(b)\kern-5pt}\includegraphics[width=0.95\linewidth]{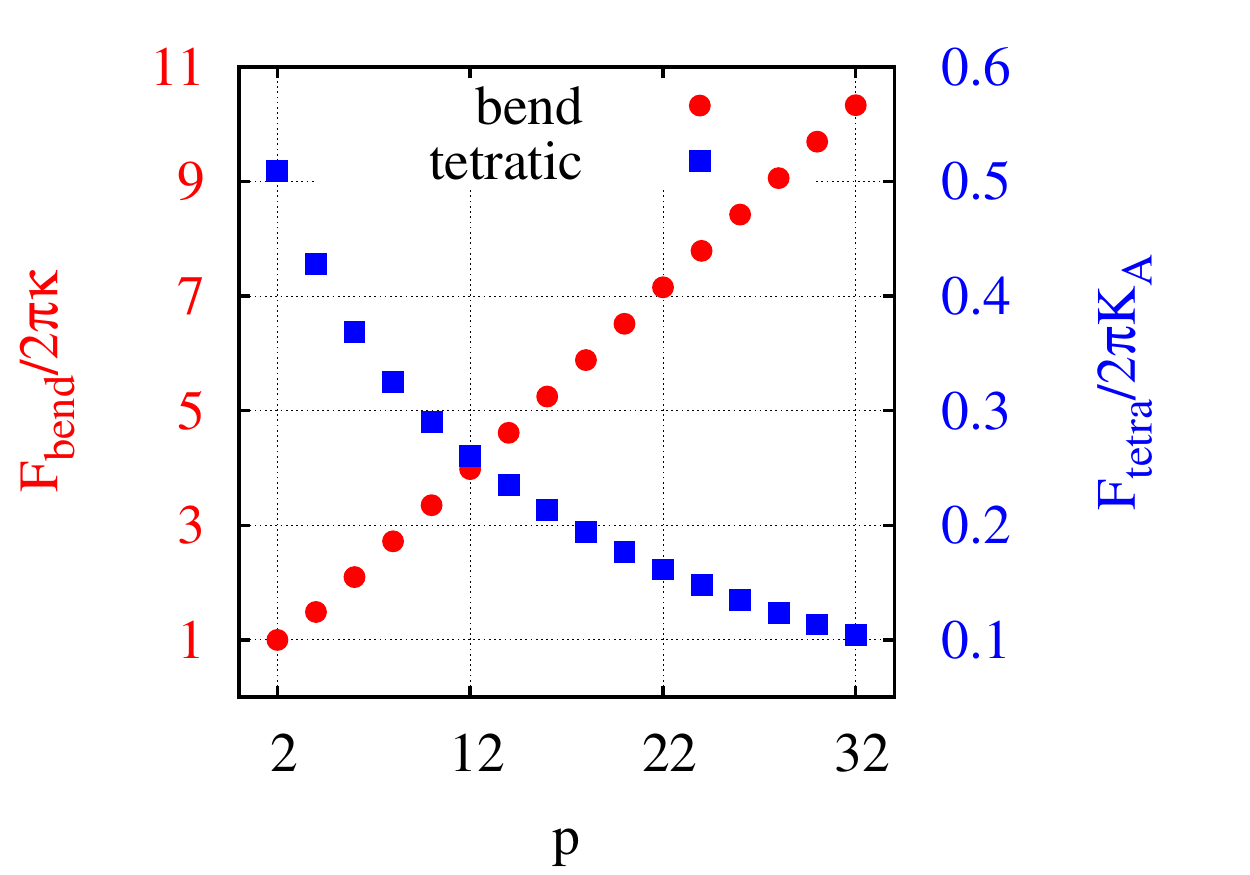}
\vskip0.1ex
\caption{\label{fig:supers}(Color online) (a) Inside: superspheroid parametrized by~\eqref{eq:param} with $p=6$. Tetratic order with  8 ($+1/4$) disclinations  at $v=\pm \pi/4$ and $u=\pi/4, 3\pi/4$ projected on the superspheroid~\cite{see:SM}. (b) The normalized tetratic (squares) and bending (circles) energies, plotted as function of the power $p$ responsible for the flattening of the superspheroid ($p=2$ is a sphere, $p\to \infty$ is a cube). We choose the angle cut-off to be $\epsilon=(\delta/a) 2^{1/p-1/2}$ with $\delta=0.01$ and $a$ determined by assuming constant area of the surface ${\cal A}=4\pi R^2$.}
\end{figure}

Gaussian curvature contributes to covariant gradients and therefore it may be preferable for highly flexible surfaces to locally flatten to minimize the tetratic free energy~\cite{bowick:review, bowick:pnas}. A good candidate for the deformed surface is a superspheroid (see Fig.~\ref{fig:supers}a), which  has the symmetry of a cube and is parametrized by 
\be\label{eq:param}
\begin{aligned}
\xi&=\frac a{(1+\tan^p v)^{1/p}}\frac 1{(1+\tan^p u)^{1/p}},\\
\eta&=\frac a{(1+\tan^p v)^{1/p}}\frac {\tan u}{(1+\tan^p u)^{1/p}},\\
\zeta&=\frac {a\tan v}{(1+\tan^p v)^{1/p}}, \quad p=2s,\ s\in {\cal N},
\end{aligned}
\ee
where $(u,v)$ are non-orthogonal curvilinear coordinates, the exponent $p$ is responsible for flattening and $a$ is the characteristic size. Thus square-shaped particles can nicely align on flattened faces at the expense of the bending energy ${\cal F}_{\rm bend}=\frac \kappa 2\int dS\,H^2$, where $H$ is the mean curvature of superspheroid concentrated at rounded edges  and $\kappa$ is the bending rigidity.
Since both bending and tetratic energies are scale invariant, it is solely the ratio $K_A/\kappa$ which determines the ground state of the system~\cite{bowick:pnas,chen:2013}. Thus a sphere ($p=2$) minimizing the bending energy will be the ground state in the limit $\kappa\gg K_A$, while deformation towards a cube ($p\to \infty$) favored by tetratic order happens when $\kappa\ll K_A$. For an intermediate ratio of $K_A$ and $\kappa$,  $K_A/\kappa\sim O(1)$, we expect a rounded cube (finite $p>2$)  as the shape minimizer of ${\cal F}_{\rm bend}+{\cal F}_{\rm tetra}$.

Within the proposed {\it ansatz}~\eqref{eq:param}, the bending energy is computed straightforwardly. It increases linearly with the exponent $p$ as shown in Fig.~\ref{fig:supers}b (circles), because of sharpening of the rounded edges of a superspheroid. The calculation of the tetratic energy involves (for details see~\cite{see:SM}) the stereographic projection of the planar configuration with eight $+1/4$ disclinations  at the vertices of two concentric squares (Fig.~\ref{fig:sphere}a) from a complex plane to a superspheroid with the coordinate transformation given by
\be\label{eq:map}
z=\frac{a \cos v\sqrt{1+\tan^2 v}e^{i\arctan(\tan u)^{p-1}}}{\sin v\sqrt{1+\tan^2v}+(1+\tan^pv)^{1/p}}.
\ee
In  Fig.~\ref{fig:supers}a we illustrate tetratic order on a superspheroid with $p=6$. Next, expressing the gradients of a vector $\bn$ within a tangent plane to a superspheroid~\eqref{eq:param}, we compute numerically the integral of tetratic free energy density $f_{\rm tetra}^{(+)}$~\eqref{eq:fpm} within one elastic constant approximation. The result is shown in Fig.~\ref{fig:supers}b (squares), where the free energy ${\cal F}_{\rm tetra}$  associated with tetratic order  decays exponentially with flattening of superspheroid ($p$). 

Our analysis suggests that the onset of shape instability from a sphere to superspheroid happens when $K_A>2\kappa$. Alternatively, one might drive this shape transformation by applying a magnetic field~\cite{oksana:prl}, which favors flat faces where self-assembled molecules can align along or orthogonal to the field. Such magnetic deformations of self-assembled nanocapsules can  be measured experimentally in reversible and controlled ways, providing a  solid test for the predictions of the model presented here and allowing one to extract the values of elastic constants $K_A$ and $\kappa$. \looseness=-1

In this Letter we have shown that 4-fold (tetratic) order on highly deformable surfaces favors the formation of cubic structures with flat faces, bending energy concentrated on edges, and disclination defects at the vertices. Cubic building blocks pack perfectly in three dimensions and provide new possibilities for self-assembly~\cite{glotzer:2007} by alignment or linking across functionalized flat faces.

\begin{acknowledgments}
The authors acknowledge financial support from the Soft Matter Program of Syracuse University.
\end{acknowledgments}


%

\vfill
\clearpage
\onecolumngrid
\renewcommand{\theequation}{\arabic{equation}$^*$}
\setcounter{equation}{0}
\renewcommand{\thefigure}{\arabic{figure}$^*$}
\setcounter{figure}{0}

\section*{Supplemental Material (SM)}

\subsection{Order parameter}

In planar geometry a unit square (an elementary constituent of tetratic phase) can be characterised by two vectors $\bn$ and $\bm$, such that $|\bn|^2=1$, $|\bm|^2=1$. However, because  of the orthogonality condition  $\bn\cdot\bm=0$ these two vectors are not independent and transform into each another under $\pi/2$ rotation:   ${\cal R}_{\pi/2}\bn=\bm$ and ${\cal R}_{\pi/2}\bm=-\bn$ or in index notations $\epsilon_{ij} n_j=m_i$ and $\epsilon_{ij} m_j=-n_i$ ($\epsilon_{12}=1$, $\epsilon_{21}=-1$, $\epsilon_{11}=\epsilon_{22}=0$). One can show that any second-rank tensor constructed from the pairs $\bn\otimes\bn$, $\bm\otimes\bm\equiv{\bf I}-\bn\otimes\bn$,  $\bn\otimes\bm$ or $\bm\otimes\bn$, or any linear combination of them is not invariant under $\pi/2$ rotation, whence a nematic-type order parameter ${\bf Q}\propto \bn\otimes\bn-{\bf I}/2$ is not a good quantity to describe tetratic phase. According to~\cite{park:singlesm}, a relevant order parameter to characterise any $k$-atic order is $\psi(\bx)=\langle \exp\{ik\alpha(\bx)\}\rangle$, $k=1,2,3\ldots$, where $\alpha(\bx)$  is an angle at point $\bx$ that vector $\bn$ makes with some fixed direction. An alternative descriptor of orientational order for tetratics ($k=4$) is  the 4-th rank symmetric traceless tensor based on vector $\bn$, written as
\begin{multline}
Q^{(4)}_{ijkl}=n_in_jn_kn_l-\frac 16 (\delta_{ij} n_kn_l +\delta_{ik}n_jn_l+\delta_{il}n_j n_k +\delta_{jk} n_in_l+\delta_{jl} n_in_k +\delta_{kl} n_in_j)+\\+\frac1{24}(\delta_{ij}\delta_{kl}+\delta_{ik}\delta_{jl}+\delta_{il}\delta_{jk}),\qquad i,j,k,l=1,2.
\end{multline}
Indeed we can show that ${\bf Q}^{(4)}$  remains invariant under $\pi/2$-rotation.

\subsection{Free energy of tetratic phase}

\begin{figure}[t]
\centering
\raisebox{50mm}{(a)\kern10pt}\includegraphics[height=5cm]{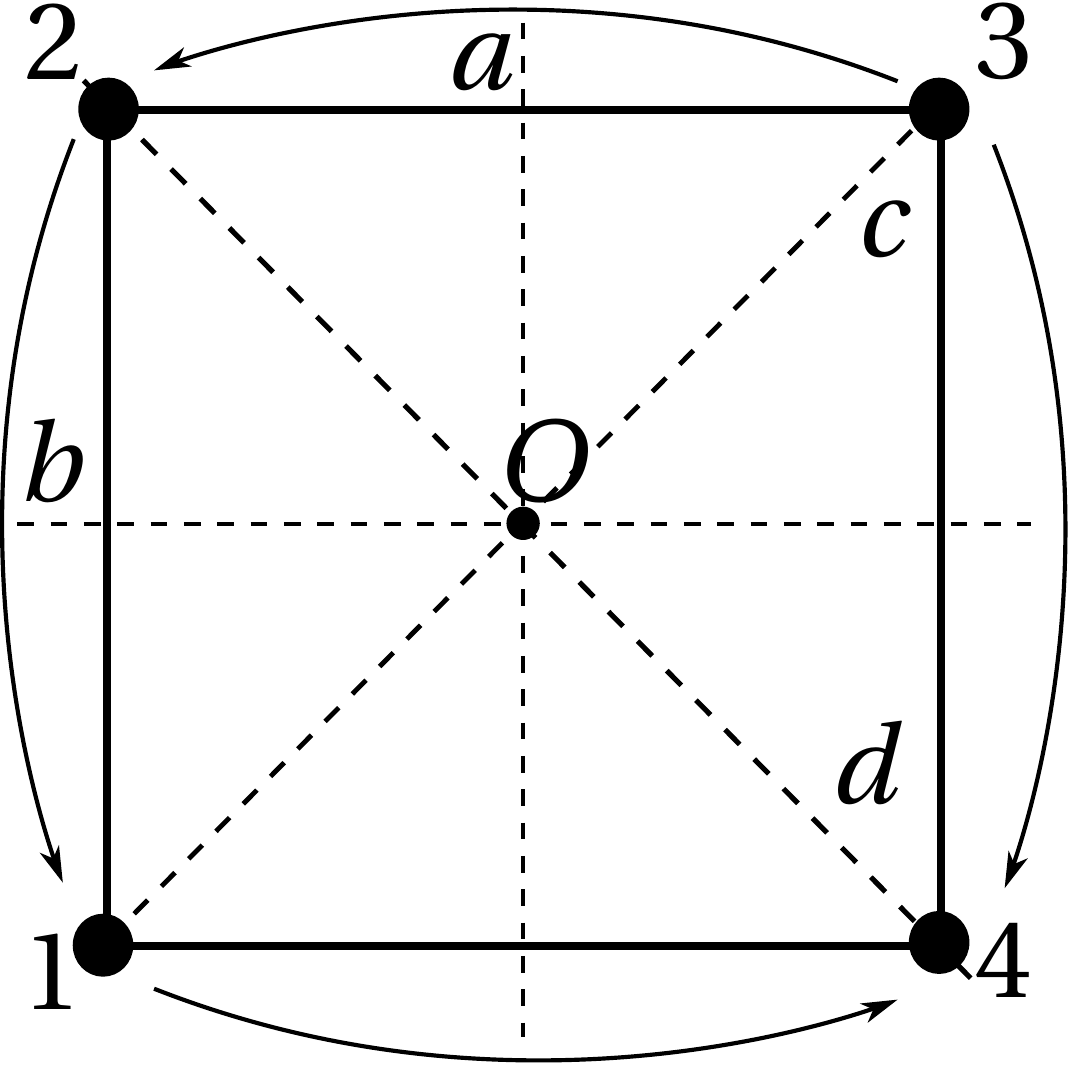}
\hfil
\raisebox{50mm}{(b)\kern10pt}\includegraphics[height=5cm]{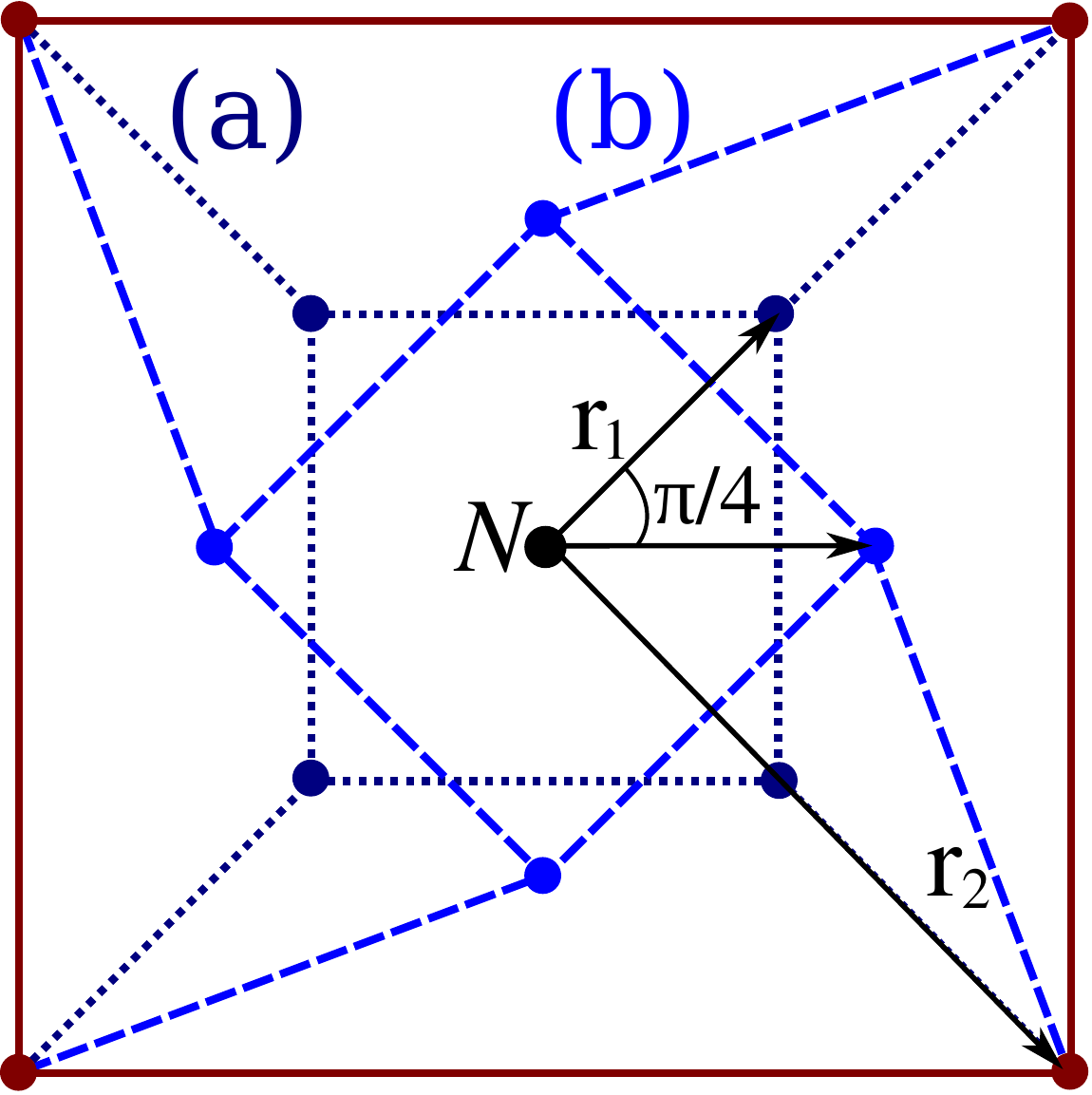}
\caption{\label{fig:square}Schematic illustration of: (a)~elements of ${\cal D}_{4h}$: $\pi/2$ rotations~\eqref{eq:rot} and mirror-reflections across line~\eqref{eq:refl}, resulting in permutations of the vertices $(1\,2\,3\,4)$ of a square and leave point $O$ fixed. (b)~two configurations of 8 defects (dots)  projected from a sphere on a plane and vice versa; analogous to Figs. 2a,b in the main text.}
\end{figure}

Similar to Oseen--Zocher--Frank free energy for nematic liquid crystals, derived algebraically (see e.g. Ref.~\cite{stewart:booksm}), we assume that the free energy density for tetratic phase $f_{\rm tetra}$ is a polynomial in $\bn$ and $\nabla\bn$, quadratic in gradients of the director  up to the lowest order
\be\label{eq:ftetra}
f_{\rm tetra}\propto \lambda_{ijklsp}\,n_i n_j \nabla_k n_l \nabla_s n_p, \qquad i,j,k,l,s,p=1,2,
\ee
which does not depend on the sign of $\bn$. Since $f_{\rm tetra}$ is an intrinsic quantity of the system it should be invariant under the operations of the dihedral point symmetry group ${\cal D}_{4h}$ of a square, so that
\be\label{eq:inv}
f_{\rm tetra}(\bn,\nabla\bn)\equiv f_{\rm tetra}({\cal R}\bn, {\cal R}\nabla\bn{\cal R}^T).
\ee
The elements of this group are in-plane $\pi/2$ rotations and mirror-reflections across line which transform a square into itself and leave its centre $O$ fixed (see Fig.~\ref{fig:square}a)
\begin{gather}\label{eq:rot}
{\cal R}_{\pi/2}=\begin{pmatrix}0 & 1 \\ -1 & 0\end{pmatrix}, \qquad {\cal R}_{\pi}=\begin{pmatrix}-1 & 0 \\ 0 & -1\end{pmatrix},\qquad {\cal R}_{3\pi/2}=\begin{pmatrix}0 & -1 \\ 1 & 0\end{pmatrix}, \qquad {\cal I}=\begin{pmatrix}1 & 0 \\ 0 & 1\end{pmatrix},\\ \label{eq:refl}
{\cal R}_{a}=\begin{pmatrix} -1 & 0 \\ 0 & 1\end{pmatrix}, \qquad {\cal R}_{b}=\begin{pmatrix} 1 & 0 \\ 0 & -1\end{pmatrix}, \qquad {\cal R}_{c}=\begin{pmatrix}0 & 1 \\ 1 & 0\end{pmatrix},\qquad  {\cal R}_{d}=\begin{pmatrix}0 & -1 \\ -1 & 0\end{pmatrix}.
\end{gather}
Note  that this representation is reducible, since the element of the group e.g. ${\cal R}_c={\cal R}_a {\cal R}_{\pi/2}$ can be obtained as a sum of two permutations. The number of independent terms in the free energy is equal to the number of irreducible representations  that transform as an invariant scalar~\cite{terentjev:1993sm} under symmetry group operations. Here we will recover the invariants of~$\lambda_{ijklsp}\,n_i n_j \nabla_k n_l \nabla_s n_p$  by directly applying the transformations~\eqref{eq:rot}, \eqref{eq:refl} and finding the energy preserving~\eqref{eq:inv} relationships between elastic constants $\lambda_{ijklsp}$.

Let us simplify our notations by introducing  small distortions of the director $\bn=n_1\bi_1+n_2\bi_2$ as 
\be\label{eq:a14}
{\cal A} =\nabla \bn=n_{i,j} \bi_i\otimes\bi_j=\begin{pmatrix}n_{1,1} & n_{1,2} \\ n_{2,1} & n_{2,2}\end{pmatrix}\equiv \begin{pmatrix}a_1 & a_2 \\ a_3 & a_4\end{pmatrix}
\ee
written  in the local orthogonal basis $\{\bi_1,\bi_2\}$. For example, under $\pi/2$ rotation
\be \label{eq:pi2}
\begin{pmatrix}n_1' \\ n_2' \end{pmatrix}={\cal R}_{\pi/2} \bn = \begin{pmatrix}n_2\\ -n_1 \end{pmatrix},\qquad
\begin{pmatrix}a_1' & a_2' \\ a_3' & a_4'\end{pmatrix}={\cal A}'={\cal R}_{\pi/2}{\cal A}{\cal R}_{\pi/2}^T=\begin{pmatrix}a_4 & -a_3 \\ -a_2 & a_1\end{pmatrix}.
\ee
Collecting  the terms $\lambda_{ij\beta\gamma }\, n_i n_j a_\beta a_\gamma=\lambda_{ij\beta\gamma}\,n_i'n_j' a_\beta'a_\gamma'$~\eqref{eq:inv} and equating the coefficients in front of $a_\beta$, $\beta=1,2,3,4$ to zero we find the relationships between elastic constants summarized as follows
\be
\begin{matrix} & n_1 a_1 & n_1 a_2 & n_1 a_3 & n_1 a_4 & n_2 a_1 & n_2 a_2 & n_2 a_3 & n_2 a_4\\[1ex]
n_1 a_1 & \lambda_{11} & 0& 0 & \lambda_{14} & 0 & \lambda_{12} & \lambda_{13} & 0\\
n_1 a_2 & & \lambda_{22} & \lambda_{23} & 0 &\lambda_{12} & 0 & 0& \lambda_{13} \\
n_1 a_3 & & & \lambda_{33} & 0&\lambda_{13} & 0 & 0& \lambda_{12} \\
n_1 a_4 & & & & \lambda_{44} & 0 & \lambda_{13} & \lambda_{12} & 0\\
n_2 a_1 & & & & & \lambda_{44} & 0 & 0& \lambda_{14}\\
n_2 a_2 & & & & & & \lambda_{33} & \lambda_{23} & 0\\ 
n_2 a_3 & & & & & & & \lambda_{22} & 0\\
n_2 a_4 & & & & & & & & \lambda_{11}
\end{matrix}
\ee
Therefore, out of 30 elastic constants $\lambda_{ij\beta\gamma }$ ($i\leftrightarrow j$, $\beta\leftrightarrow \gamma$) only 8 enter the free energy~\eqref{eq:ftetra}. 
\begin{multline}\label{eq:f8}
f_{\rm tetra} \propto \lambda_{11}\underbrace{(n_1^2 a_1^2 +n_2^2 a_4 ^2)}_{\iota_1}+\lambda_{22}\underbrace{(n_1^2 a_2^2 +n_2^2 a_3 ^2)}_{\iota_1} +\lambda_{33}\underbrace{(n_1^2a_3^2+n_2^2 a_2 ^2)}_{\iota_2}+\lambda_{44}\underbrace{(n_1^2a_4^2+n_2^2 a_1 ^2)}_{\iota_3}+\\[1ex] 4\lambda_{13}\underbrace{n_1 n_2 (a_1 a _3 +a_2 a_4)}_{-\iota_1} +2\lambda_{23}(n_1^2+n_2^2)\underbrace{a_2 a_3}_{\iota_4}+2\lambda_{14}(n_1^2 + n_2^2) \underbrace{a_1 a_4}_{\iota_4}+4\lambda_{12}\underbrace{n_1 n_2 (a_1 a_2+a_3 a_4)}_{-\iota_4}.
\end{multline}
Moreover, this algebraic expression can be further simplified accounting for the normalization of the director $n_in_i=1$,  $n_i\p_jn_i=0$ and 4 independent invariants:
\begin{gather}
n_1^2+n_2^2=1,\qquad n_1 a_1+n_2 a_3=0,\qquad n_1  a_2+n_2 a_4=0,\\
\iota_1=n_1^2a_1^2+n_2^2a_4^2=n_1^2a_3^2+n_2^2a_3^2=-n_1n_2(a_1a_3+a_2a_4), \\
\iota_2=n_1^2a_3^2+n_2^2 a_2 ^2, \quad \iota_3=n_1^2a_4^2+n_2^2 a_1 ^2, \\
\iota_4=a_2a_3= a_1 a_4=-n_1n_2(a_1a_2+a_3a_4).
\end{gather}
Here we have also assumed that the saddle-splay term~\cite{napoli:2012sm} does not contribute to the bulk free energy
\be\label{eq:det}
2(a_1a_4-a_2a_3)=2\det(\nabla\bn)\equiv[\tr(\nabla\bn)]^2-\tr[(\nabla\bn)^2]=\nabla\cdot\big[(\nabla\cdot\bn)-(\bn\cdot\nabla)\bn\big]=0.
\ee
Combining the terms in~\eqref{eq:f8} we get
\begin{multline}\label{eq:f4}
f_{\rm tetra} \propto \lambda_{33}(n_1^2a_3^2+n_2^2 a_2 ^2)+\lambda_{44}(n_1^2a_4^2+n_2^2 a_1 ^2)+(4\lambda_{13}-\lambda_{11} -\lambda_{22})n_1 n_2 (a_1 a _3 +a_2 a_4)  +\\ +2\underbrace{(\lambda_{23} +\lambda_{14}-2\lambda_{12})}_{=0} a_1 a_4.
\end{multline}
In absence of information about the values of phenomenological constants $\lambda_{ij}$, we neglect the last term, assuming $\lambda_{12}=\lambda_{23}=\lambda_{24}$. Otherwise, this term will contribute to the positive definiteness of the quadratic form~\eqref{eq:f4}.

Next,  we consider two limiting cases, namely
\begin{itemize}
\item the one constant approximation $\lambda_{11}=\lambda_{22}=\lambda_{33}=\lambda_{44}=\lambda_{13}=1$, yielding~\eqref{eq:f4} as
\be\label{eq:fplu}
f_{\rm tetra}^{(+)} =\frac{K_A}2\big[ (n_1 a_3+n_2 a_1)^2+(n_1 a_4+n_2 a_2)^2\big]=\frac{K_A}2\sum_{\substack {i=1\\i\neq j}}^2\big (n_i (\nabla\bn)_{ji}+n_j(\nabla\bn)_{ii}\big)^2,
\ee
where $K_A$ is phenomenological elastic constant.
\item the vanishing of off-diagonal term $\lambda_{11}=\lambda_{22}=\lambda_{33}=\lambda_{44}=1$ and $\lambda_{13}=0$
\be\label{eq:min}
f_{\rm tetra}^{(-)} =\frac{K_A}2\big[ (n_1 a_3-n_2 a_1)^2+(n_1 a_4-n_2 a_2)^2\big]=\frac{K_A}2\sum_{\substack {i=1\\i\neq j}}^2\big (n_i (\nabla\bn)_{ji}-n_j(\nabla\bn)_{ii}\big)^2.
\ee
\end{itemize}
Both cases guarantee that the free energy $f_{\rm tetra}^\pm$ is positive definite if $K_A>0$. Parametrizing the director in a local orthonormal system of coordinates as $\bn=\cos\alpha\,\bi_1+\sin\alpha\,\bi_2$ we get the tetratic energy density as
\be\label{eq:ftpm}
f_{\rm tetra}^{(\pm)}=\frac {K_A}2\left\{ \begin{aligned} \cos^2(2\alpha)|\nabla\bn|^2, \quad &\lambda_{ij}=1,\\ |\nabla\bn|^2,\quad &\lambda_{ii}=1, \lambda_{13}=0, \end{aligned}\right.
\ee
which we distinguish below as tetratic model$^{(+)}$ and  tetratic model$^{(-)}$.

To  analyse the stability of the tetratic energy~\eqref{eq:f4}, let us rewrite it for simplicity in the Cartesian coordinates,
\begin{multline}
f_{\rm tetra}= \lambda_{33}\big(\cos^4\alpha \alpha_{,x}^2+\sin^4\alpha \alpha_{,y}^2\big) +\lambda_{44}\big(\sin^4\alpha \alpha_{,x}^2+\cos^4\alpha \alpha_{,y}^2\big)+\\+ (\lambda_{11} +\lambda_{22}-4\lambda_{13})\sin^2\alpha\cos^2\alpha\big(\alpha_{,x}^2+\alpha_{,y}^2\big).
\end{multline}
This quadratic form is positive definite if and only if  the following inequality holds
\be
 (\lambda_{11} +\lambda_{22}-4\lambda_{13})^2\leqslant 4 \lambda_{33}\lambda_{44}\quad \xrightarrow{\lambda_{ii}=1}\quad (1-2\lambda_{13})^2\leqslant 1
\ee
Both cases $\lambda_{13}=1$ (one-constant approximation) and $\lambda_{13}=0$, mentioned above, are similar with respect to the stability of the tetratic free energy. Reducing the number of phenomenological elastic constants allows to study solutions of the problems, formulated below, analytically.

\subsection{The Euler--Lagrange equations on a plane}

In a planar geometry the director can be parametrized as $\bn=\cos\alpha(x,y)\bi_x+\sin\alpha(x,y)\bi_y$ and  the tetratic free energy is written as
\be\label{eq:fplan}
{\cal F}_{\rm tetra}^{(+)}=\frac{K_A}2\iint dx\,dy\,\cos^2(2\alpha)\big[(\p_x\alpha)^2+(\p_y\alpha)^2\big].
\ee
One can either i)~notice immediately that the replacement of variable $\sin(2\alpha)=\beta/2$ brings $f_{\rm tetra}^{(+)}=(\p_x\beta)^2+(\p_y\beta)^2$ to the tetratic model$^{(-)}$ with known solution $\beta=\arctan(y/x)$ or ii)~study the behavior of the following Euler--Lagrange equation associated with~\eqref{eq:fplan} as  
\be
\cos(2\alpha)\big[\p_{xx}\alpha+\p_{yy}\alpha\big] -2\sin(2\alpha)\big[(\p_x\alpha)^2+(\p_y\alpha)^2\big]=0.
\ee
To solve this second-order nonlinear  PDE we assume that angle $\alpha(x,y)$ depends solely on the ratio $\xi\equiv y/x$ and not on the absolute distance. Then we get 
\be
\cos(2\alpha)(1+\xi^2)\p_{\xi\xi} \alpha-2\sin(2\alpha)(1+\xi^2)(\p_\xi\alpha)^2+2\xi\cos(2\alpha)\p_\xi\alpha=0,
\ee
whose solution can be written in the closed form as
\be\label{eq:arcsinsm}
\alpha(\vp)=\frac 12 \arcsin\bigg\{\frac{\vp}\pi\bigg\}, \qquad \vp=\arctan\frac y x.
\ee
The integration constant at the denominator is chosen by requiring the topological charge of the defect 
at the point $(0,0)$ to be~\cite{bowick:reviewsm}
\be\label{eq:charge}
\frac 1{2\pi}\oint_\gamma d\alpha=+\frac 14.
\ee

\subsection{Tetratic energy on  a sphere}

Let ${\cal S}$ be a two-dimensional surface in ${\mathbb R}^3$, parametrized by $\bx$ as function of curvilinear coordinate $(u,v)$ and the metric of the surface ${\cal S}$ as
\be\label{eq:ds}
ds^2=E\,du^2+2F \, du\,dv+G\, dv^2, \qquad 
\ee
where $E=\p_u\bx\cdot\p_u\bx$, $F=\p_u\bx\cdot\p_v\bx$, and $G=\p_v\bx\cdot\p_v \bx$ are the coefficients of the first fundamental form. The orthogonal parametrization assumes $F\equiv 0$, so the pair of orthonormal basis vectors can be chosen as  $\bi_1=\p_u\bx/\sqrt{E}$, $\bi_2=\p_v\bx/\sqrt{G}$. Then the covariant derivative of the director $\bn=\cos\alpha\,\bi_1+\sin\alpha\,\bi_2$  is~\cite{napoli:2012sm,bowick:reviewsm,docarmosm}
\begin{subequations}
\begin{gather}
\nabla\bn=-\sin\alpha\,\bi_1\otimes\nabla\alpha+\cos\alpha\,\nabla\bi_1+\cos\alpha\,\bi_2\otimes\nabla\alpha+\sin\alpha\,\nabla\bi_2,\label{eq:dn}\\
\nabla\alpha=\vartheta_1\,\bi_1+\vartheta_2\,\bi_2,\qquad \vartheta_1=\frac{\p_u\alpha}{\sqrt E},\quad \vartheta_2=\frac{\p_v\alpha}{\sqrt G}\label{eq:ntheta}\\
\nabla\bi_1=\kappa_1\,\bi_2\otimes\bi_1+\kappa_2\,\bi_2\otimes\bi_2,\qquad \kappa_1=-\frac{\p_v E}{2E\sqrt G},\label{eq:ne1}\\
\nabla\bi_2=-\kappa_1\,\bi_1\otimes\bi_1-\kappa_2\,\bi_1\otimes\bi_2,\qquad \kappa_2=\frac{\p_uG}{2G\sqrt E},\label{eq:ne2}
\end{gather}
\end{subequations}
where $\kappa_1$ and $\kappa_2$ are the geodesic curvatures of the lines of curvature on $\cal S$. In the following section we derive the general expression for the covariant derivative of a scalar and of a vector within non-orthogonal parametrization. The final result within the orthogonal curvilinear coordinates $(u,v)$ for $\nabla\bn$ reads
\begin{multline}
\nabla\bn=-\sin\alpha(\vartheta_1+\kappa_1)\bi_1\otimes\bi_1 -\sin\alpha(\vartheta_2+\kappa_2)\bi_1\otimes\bi_2+\\+\cos\alpha(\vartheta_1+\kappa_1)\bi_2\otimes\bi_1+\cos\alpha(\vartheta_2+\kappa_2)\bi_2\otimes\bi_2,
\end{multline}
and thus the tetratic free energy~\eqref{eq:ftpm} is
\be
{\cal F}_{\rm tetra}^{(\pm)}=\frac{K_A}2\iint du\,dv\,\sqrt{EG}\, \big[(\vartheta_1+\kappa_1)^2+(\vartheta_2+\kappa_2)^2\big]\left\{ \begin{aligned} \cos^2(2\alpha), \quad \mbox{model}^{(+)},\\ 1, \quad \mbox{model}^{(-)}. \end{aligned}\right.
\ee

For a sphere of radius $R$: ${\bf r}(u,v)=(R \sin v\cos u,R\sin v\sin u,R\cos v)$ we get the metric $ds^2=R^2\sin^2v \,(du)^2+R^2(dv)^2$, so that the tetratic free energy in the one constant approximation
\be\label{eq:sphere}
{\cal F}_{\rm sph}^{(+)}=\frac{K_A}2\iint du\,dv\,\sin v\, \cos^2(2\alpha) \bigg[(\p_v\alpha)^2+\frac 1{\sin^2v} (\cos v-\p_u\alpha)^2 \bigg].
\ee
Solving the Euler--Lagrange equation analytically in spherical coordinates  seems cumbersome. However, after projecting stereographically on a plane with polar coordinates, so that we substitute $r=R\tan\frac v2$ and  $u\mapsto\vp$ in~\eqref{eq:sphere}, we get
\be\label{eq:projsm}
{\cal F}_{\rm proj}^{(\pm)}=\frac{K_A}2 \int_0^\infty r\,dr\int_0^{2\pi}d\vp\,\bigg[(\p_r\alpha)^2 + \frac 1{r^2} \bigg(\frac{R^2-r^2}{R^2+r^2} -\p_\vp\alpha\bigg)^2\bigg]\left\{ \begin{aligned} \cos^2(2\alpha),\\ 1. \end{aligned}\right.
\ee
If we assume no radial dependence of $\alpha$ then the solution to the Euler--Lagrange equation for the projected sphere coincides with~\eqref{eq:arcsinsm}.

\subsection{Geodesics and model$^{(\pm)}$ on a sphere}

Conventionally, it is assumed that the lowest energy configuration corresponds to the furthest separation of defects. As shown in the main text, this is indeed the case for model$^{(-)}$. In general, the free energy of $k$-atic on a sphere can be rewritten as the sum of all pair interactions between defects $i$ and $j$~\cite{nelson:2002sm,bowick:reviewsm}
\be
{\cal F}_{\rm pair}=-\frac {\pi K_A}{2k^2}\sum_{i\neq j} q_i q_j \log\bigg(\frac{1-\cos\beta_{ij}}2\bigg) +E(R) \sum_j q_j^2,
\ee
which in turn depends on geodesic distance on a sphere between two defects $\cos\beta_{ij}=\cos v_i\cos v_j+\sin v_i \sin v_j \cos(u_i-u_j)$, $q_i$ and $q_j$ are the winding numbers, and $E(R)$ is the self-energy of defects. The straightforward calculation of  ${\cal F}_{\rm pair}$ within tetratic model ($k=4$) for i) 8 defects at the vertices defining inscribed cube with $v_i=\{\pi/4, 3\pi/4\}$ and $u_i=(2m-1)\pi/4$, $m=1,2,3,4$ gives ${\cal F}_{\rm cube}=3.7183K_A$, while ii) for 8 defects at the vertices defining inscribed twisted cube with $v_i=\{\pi/4,3\pi/4\}$ and $u_j=\{(2m-1)\pi/4,m \pi/2\}$  gives ${\cal F}_{\rm twist}=3.7156 K_A<{\cal F}_{\rm cube}$ (listed in Table I in the main text). 

\begin{figure}[th]
\centering
\raisebox{45mm}{(a)}\includegraphics[height=45mm]{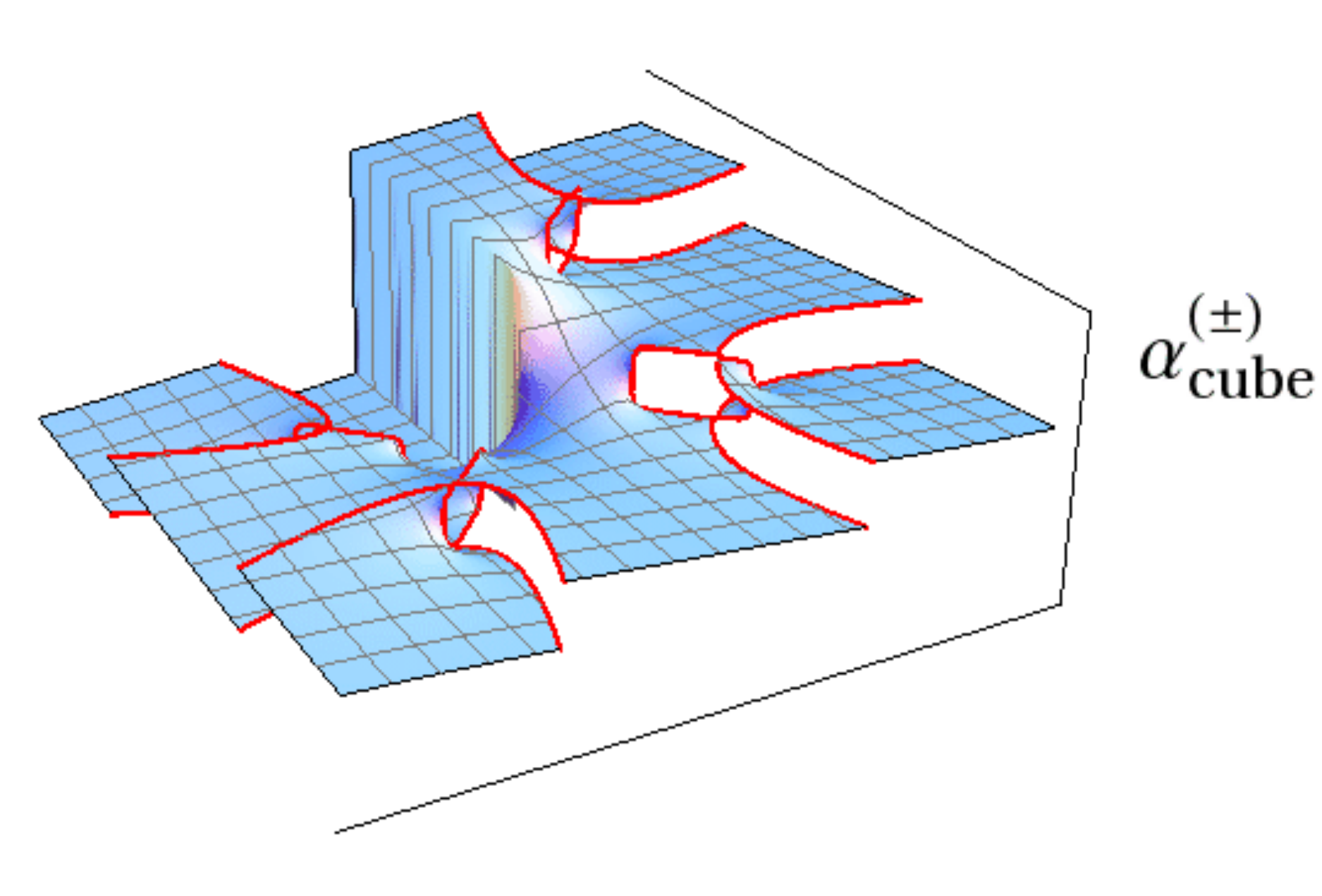}
\hfil
\raisebox{45mm}{(b)}\includegraphics[height=45mm]{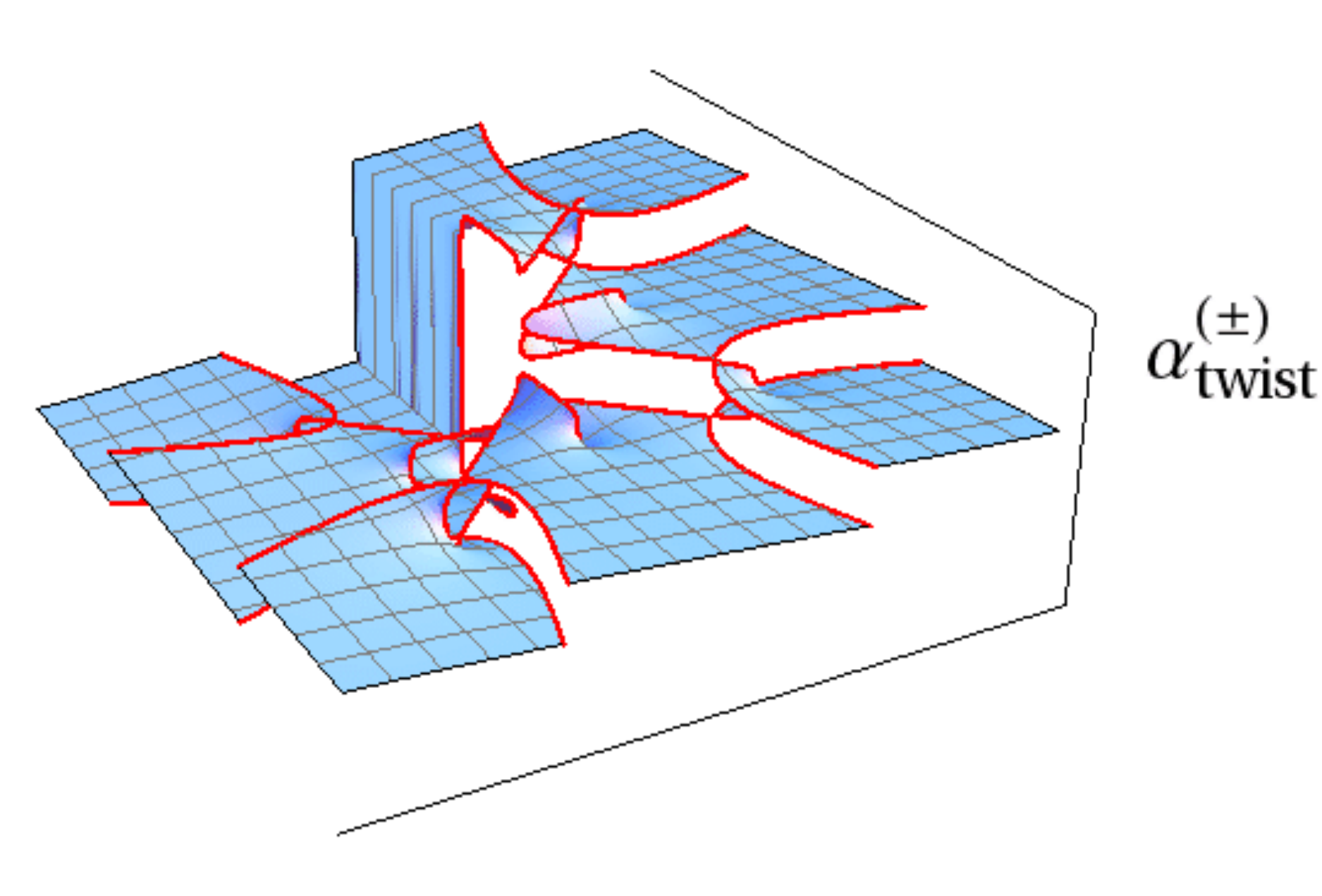}
\caption{\label{fig:riemann} Visualization of the Riemann surface for $\alpha^{(\pm)}$ with branch points corresponding to~\eqref{eq:defab}, (a) and (b), respectively; at branch cut the angle $\alpha^{(\pm)}$ jumps by $\pi/2$.}
\end{figure}

In Fig.~\ref{fig:square}b we show schematically the stereographic projection of these two defect configurations from a sphere on a plane (see also Fig. 2a,b in the main text) with 
\be\label{eq:defab}
\mbox{(a)}\ r_m= r_{1,2} e^{i\pi(2 m-1)/4}, \quad\mbox{or}\quad \mbox{(b)}\ r_m= r_1 e^{i\pi(2 m-1)/4},\ r_m= r_2 e^{i\pi m/2},\quad m=1,2,3,4,  
\ee
where $r_1=R \tan \frac{v_1} 2=R\tan \frac\pi 8$ and $r_2=R \tan\frac{v_2}2=R\tan \frac{3\pi} 8$.
Within the models$^{(\pm)}$ the angle $\alpha$ is given by
\begin{align}
\alpha^{(-)}(z)&=\frac \pi2-\frac 14 \Im\log\{(z^4\pm r_1^4)(z^4+r_2^4)\}+\Im \log(z),\qquad z=x+i y=re^{i\vp},\\
\alpha^{(+)}(z)&=\frac \pi2-\frac 12 \arcsin\bigg\{\frac{\Im\log\{(z^4\pm r_1^4)(z^4+r_2^4)\}}\pi\bigg\} +\Im \log(z), 
\end{align}
where $+$ sign in the denominator corresponds to straight cube (a) and $-$ sign to a twisted cube (b). The origin of defect of charge $+1$ at the North ($N$) pole (projected to $r=0$) comes from the change of the variables $(r,\vp)$ to $(x,y)$. 
Note that $\arcsin(z)=-i\log(iz+\sqrt{1-z^2})$ and representation of complex functions in terms of Riemann surfaces is similar. In Fig.~\ref{fig:riemann} we show these multivalued functions; when crossing the branch cut the value of $\alpha^{(\pm)}$ changes by $\pi/2$.

\subsection{Stereographic projection of a superspheroid}

\begin{figure}[th]
\centering
\raisebox{55mm}{(a)\kern-15pt}\includegraphics[height=55mm]{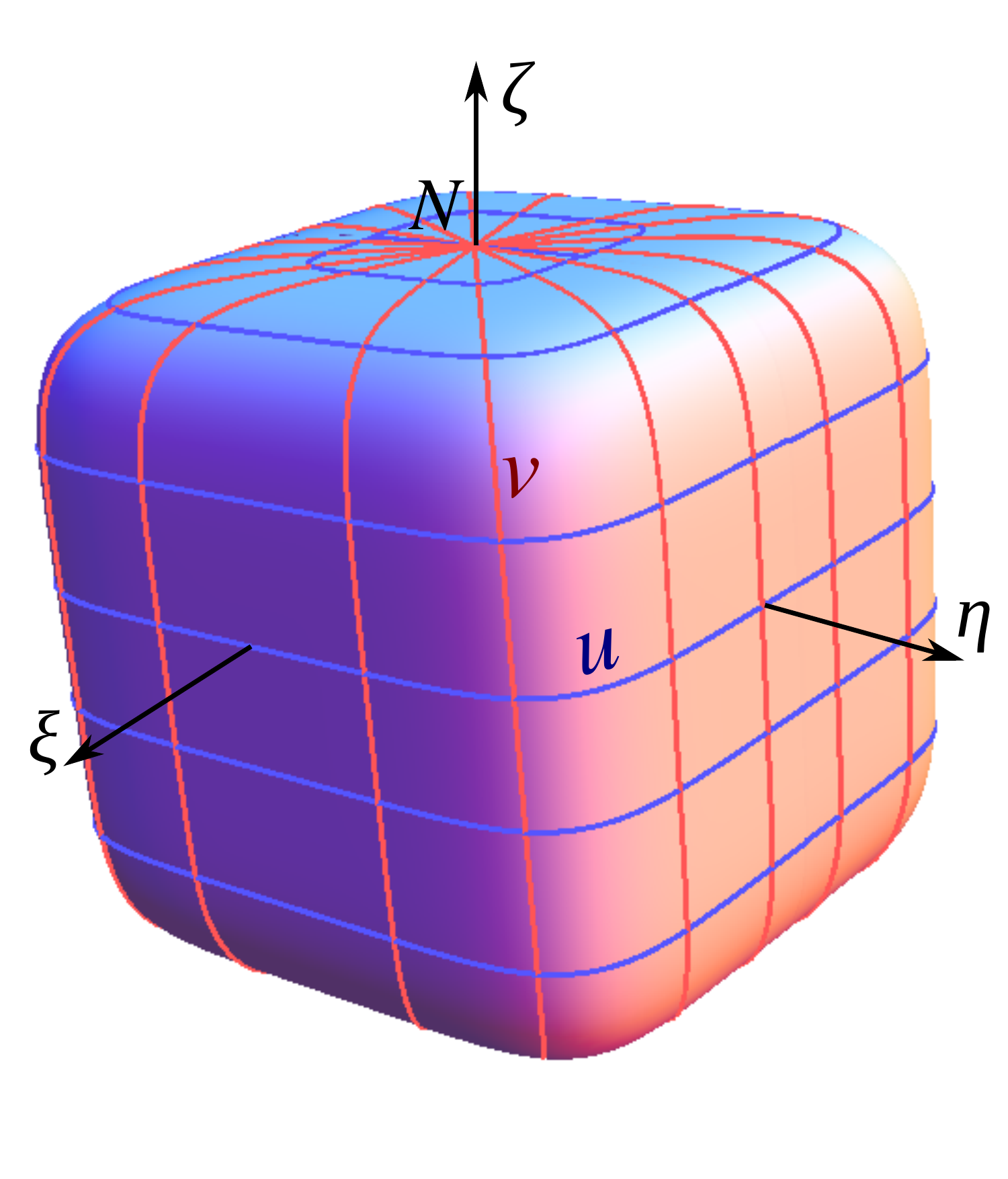}
\hfil
\raisebox{55mm}{(b)\kern-15pt}\includegraphics[height=55mm]{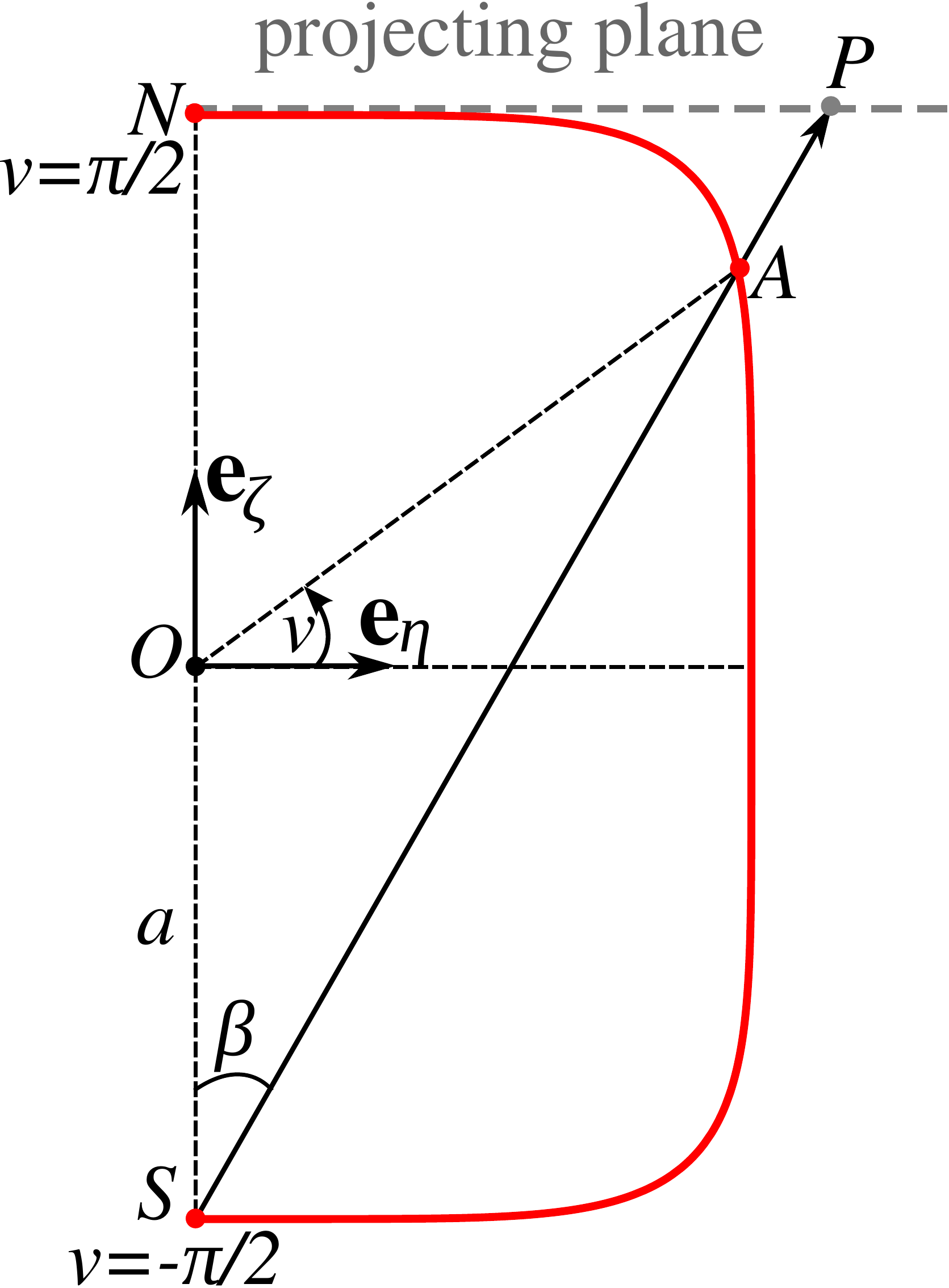}
\hfil
\raisebox{55mm}{(c)\kern-15pt}\includegraphics[height=55mm]{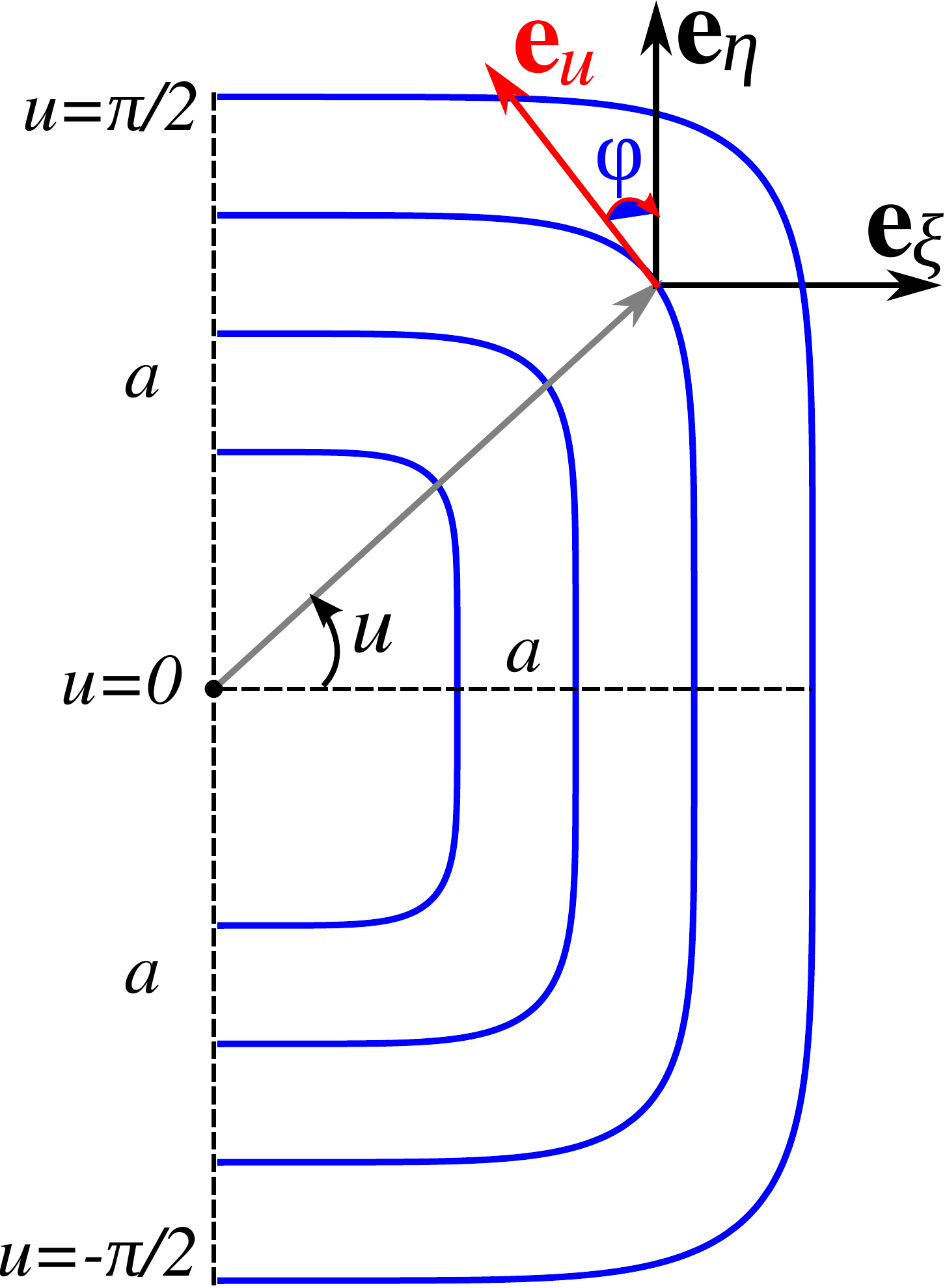}
\caption{\label{fig:proj} (a) Superspheroid~\eqref{eq:super} for $p=6$; (b)  its projection on a plane; (c) winding number associated with superellipse.}
\end{figure}

In this section we derive the coordinates transformation while projecting superspheroid (Eq.~(6) main text, Fig.~\ref{fig:proj}a) stereographically on a tangent plane at the North pole ($N$). In Fig.~\ref{fig:proj}b,c we show the cross sections of superspheroid. To find the distance $NP$, where $P$ is the point of intersection between the ray from the South pole ($S$) and the projecting plane, we consider the triangle $SOA$ (see Fig.~\ref{fig:proj}b) and apply the laws of cosines and sines
$$
\Delta SOA:\qquad \left\{ \begin{gathered} SA^2=SO^2+OA^2-2 SO\cdot OA\cos(\pi/2+v)\\
\frac{SA}{\sin(\pi/2+v)}=\frac{OA}{\sin\beta}\quad \to \quad \quad \tan^2\beta=\frac{OA^2 \sin^2(\pi/2+v)}{SA^2-OA^2\sin^2(\pi/2+v)}
\end{gathered}\right.
$$
From our parametrization by the curvilinear coordinate $v$ it follows that  $OA=\sqrt{\eta^2+\zeta^2}$ where $\eta=a (1+\tan^p v)^{-1/p}$, $\zeta=a \tan v(1+\tan^p v)^{-1/p}$. Substituting this result into $\tan \alpha$ we get
\be
\tan\beta=\frac{\cos v\sqrt{1+\tan^2 v}}{(1+\tan^p v)^{1/p}+\sin v\sqrt{1+\tan^2v}},
\ee
and whence the projected distance $NP=2a \tan\beta$ (see Eq.~(7) in the main text).

In Fig.~\ref{fig:proj}c we show the orthogonal direction $(\xi,\eta)$ of the projecting plane.  The tangent unit vector to superellipses is
\be
\bi_u=\frac {\p_u\xi}{\sqrt{(\p_u\xi)^2+(\p_u\eta)^2}} \bi_\xi+\frac {\p_u\eta}{\sqrt{(\p_u\xi)^2+(\p_u\eta)^2}} \bi_\eta,
\ee
where  $\xi=a (1+\tan^p u)^{-1/p}$, $\eta=a \tan u(1+\tan^p u)^{-1/p}$. The angle between $\bi_u$ and $\eta$-axis is 
\be
\tan\vp=-\frac{\p_u\xi}{\p_u\eta}=(\tan u)^{p-1}.
\ee
The winding number associated with the tangent vector field $\bi_u$ is
\be
s=\frac 1{2\pi}\int_{-\pi/2}^{\pi/2} du\,(\bi_u\times\p_u\bi_u)=\frac{\arctan\{(\tan u)^{-1+p}\}}{2\pi}=\frac 12,
\ee
which is exactly what one expects for half a circle, or half  a square, or anything in between.

\subsection{Non-orthogonal parametrization}

The superspheroid can be parametrized as (see Eq.~(6) in the main text)
\be\label{eq:super}
\bx=\bigg(\frac{a}{(1 + \tan^p v)^{1/p}(1 + \tan^p u)^{1/p}},\frac{a \tan u}{(1 + \tan^p v)^{1/p}(1 + \tan^p u)^{1/p}},\frac{a\tan v}{(1 + \tan^p v)^{1/p}}\bigg),
\ee
where $a$ is the characteristic size and the exponent $p\geqslant 2$.
Then the metric~\eqref{eq:ds} for superspheroid contains the following elements
\begin{subequations}\label{eq:EFG}
\begin{align}
E&=\frac{a^2}{\sin^2 u \cos^2 u} (1+\tan^p u)^{-2 (1+p)/p} (\tan^2 u+ \tan^{2 p}u ) (1+\tan^p v)^{-2/p},\\
F&=\frac{a^2(1+\tan^p u)^{-(2+p)/p}}{\sin u\sin v \cos u\cos v}  (\tan^p u+\tan^2 u)\tan^p v (1+\tan^p v)^{-(2+p)/p},\\
G&=\frac{a^2(1+\tan^p u)^{-2/p}}{\sin^2 v \cos^2 v} (1+\tan^p v)^{-2 (1+p)/p} \bigg((1+\tan^p u)^{2/p} \tan^2 v+\frac{\tan^{2p} v}{\cos^2 u} \bigg).
\end{align}
\end{subequations}

\begin{figure}[th]
\centering
\raisebox{32mm}{(a)\kern-10pt}\includegraphics[height=30mm]{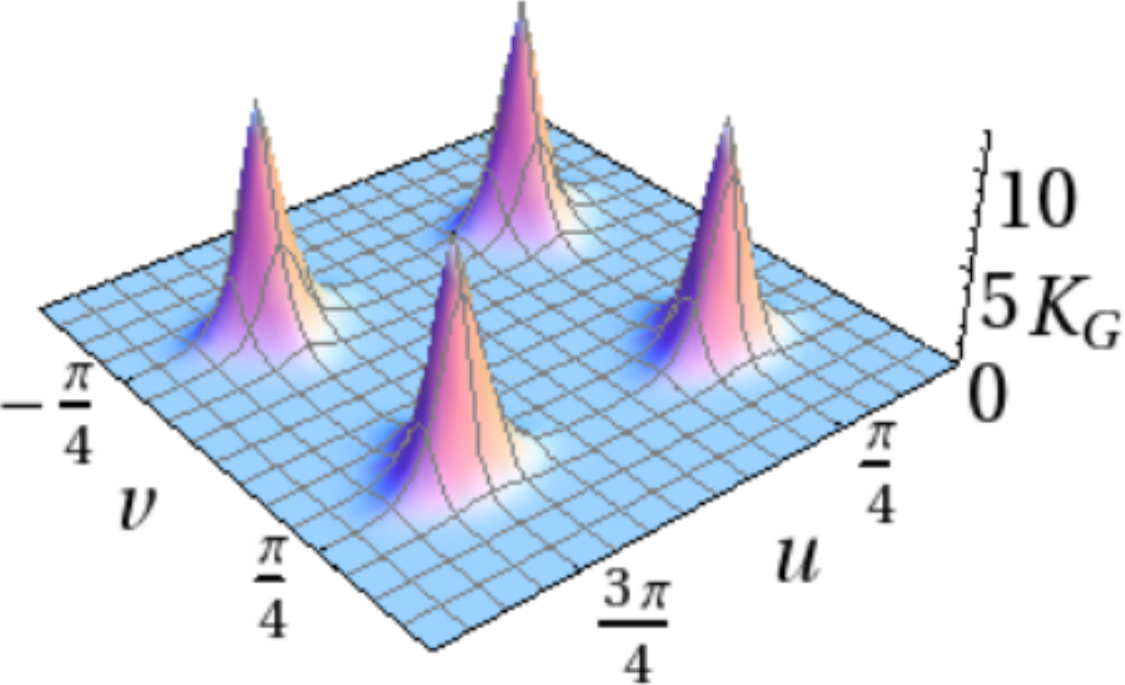}
\hfil
\raisebox{32mm}{(b)\kern-10pt}\includegraphics[height=30mm]{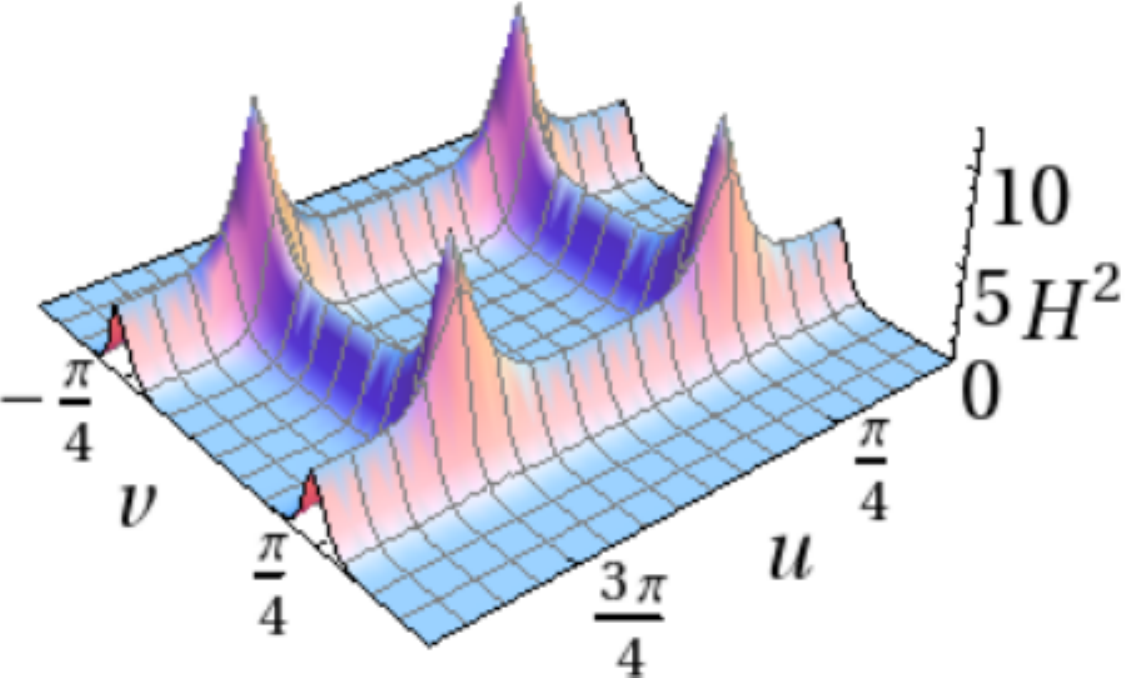}
\hfil
\raisebox{32mm}{(c)\kern-10pt}\includegraphics[height=30mm]{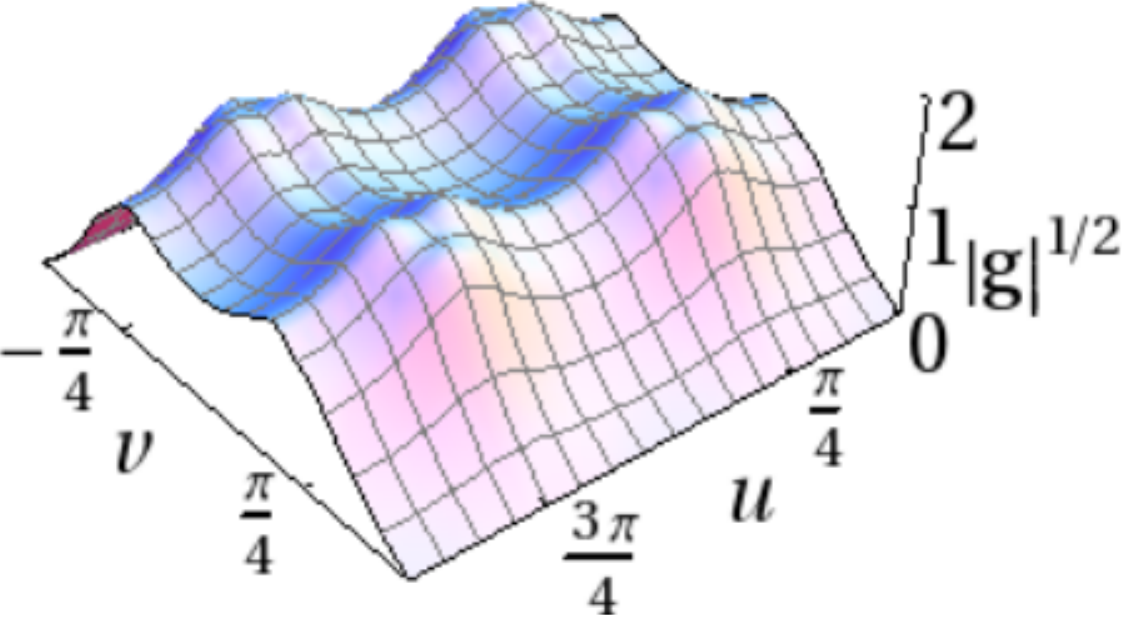}
\caption{\label{fig:curv} The Gaussian curvature (a), the mean curvature squared (b) and the metric $\sqrt{EG-F^2}$ for superspheroid with $a=1$ and  $p=6$ (as in Fig.~3a).}
\end{figure}

Because  $F\neq0$ we need to generalize the covariant derivatives to non-orthogonal parametrization. The pair of orthonormal basis vectors $\{\bi_1,\bi_2\}$ can be chosen as
\be\label{eq:basis}
\bi_1= \frac{\bx_{,u}}{\sqrt E},\qquad \bi_2=\frac{-F/E \bx_{,u}+\bx_{,v}}{\sqrt{G-F^2/E}},
\ee
where  comma indicates the partial derivative, so $ _{,u}\equiv\p_u$ and $_{,v}\equiv\p_v$.

In the following we will use the notations of D. A. Clarke from ``A primer on Tensor Calculus'', referred as~\cite{clarkesm}. The metric $g_{ij}$  is given by
\be\label{eq:gij}
g_{ij}=h_{(i)}h_{(j)} \hat \bi_{(i)}\cdot \hat \bi_{(j)}, \qquad i,j=1,2.
\ee
Here $\hat e_{(i)}$ are the physical unit vectors and $h_{(i)}$ are the scale factors. The parentheses are used to prevent the summation rule. Comparing with~\eqref{eq:ds}, we have
\be\label{eq:scale}
\hat \bi_{(1)}=\frac{\bx_{,u}}{h_{(1)}},\quad \hat \bi_{(2)}= \frac{\bx_{,v}}{h_{(2)}},\qquad h_{(1)}\equiv\sqrt{E},\quad h_{(2)}\equiv\sqrt{G},\qquad \hat \bi_{(1)}\cdot \hat \bi_{(2)}\equiv \frac{F}{\sqrt{EG}}.
\ee
For any vector $\bA=\sum_i A_{(i)}\hat \bi_{(i)}$, where $A_{(i)}$ are the physical components with respect to the basis $\hat \bi_{(i)}$ which can be related to the contravariant components as 
\be
A_{(i)}=h_{(i)}A^i=h_{(i)}g^{ij}A_j\quad \to \quad (\nabla \alpha)_{(i)}=h_{(i)}g^{ij} \p_j \alpha\, \hat\bi_{(i)}.
\ee
Substituting
\be
g^{ij}=\frac1{EG-F^2}\begin{pmatrix} G&-F\\-F&E\end{pmatrix}
\ee
we find the same result as~\cite{docarmosm} (p.~102 Ex.14), namely
\be
\nabla \alpha=\frac{G\alpha_{,u}-F\alpha_{,v}}{EG-F^2} \bx_{,u} + \frac{-F\alpha_{,u}+E\alpha_{,v}}{EG-F^2} \bx_{,v}.
\ee
Changing to orthonormal basis~\eqref{eq:basis}: $\bx_{,u}=\bi_1\sqrt{E}$ and $\bx_{,v}=\bi_1F/\sqrt{E}+\bi_2\sqrt{(EG-F^2)/E}$, we get
\be\label{eq:boxtheta}
\boxed{\nabla\alpha=\frac{\alpha_{,u}}{\sqrt E}\bi_1+\frac{E\alpha_{,v}-F\alpha_{,u}}{\sqrt{E(EG-F^2)}}\bi_2}.
\ee
One can check that for $F\equiv0$ the last expression falls into~\eqref{eq:ntheta}.

Within the same framework  we can compute the covariant derivative of basis vectors $\{\bi_1,\bi_2\}$, entering~\eqref{eq:dn}. The covariant derivative of the covariant vector is 
\be\label{eq:nA}
\nabla_i A_j=\p_iA_j-\Gamma_{ij}^k A_k\equiv{\cal G}_{ij}.
\ee
Here $\Gamma_{ij}^k$ are the Christoffel symbols given in terms of the partial derivatives of the metric tensor~\eqref{eq:gij} by 
\be
\Gamma_{ij}^k=\frac{g^{kl}}2\big(g_{jl,i}+g_{li,j}-g_{ij,l} \big),
\ee
or written explicitly
\begin{subequations}
\begin{gather}
\Gamma_{11}^1=\frac{GE_{,u}-F(2F_{,u}-E_{,v})}{2(EG-F^2)},\qquad
\Gamma_{11}^2=\frac{FE_{,u}+E(2F_{,u}-E_{,v})}{2(EG-F^2)},\\
\Gamma_{12}^1=\Gamma_{21}^1=\frac{GE_{,v}-FG_{,u}}{2(EG-F^2)},\qquad
\Gamma_{12}^2=\Gamma_{21}^2=\frac{EG_{,u}-FE_{,v}}{2(EG-F^2)},\\
\Gamma_{22}^1=\frac{-FG_{,v}+G(2F_{,v}-G_{,u})}{2(EG-F^2)},\qquad
\Gamma_{22}^2=\frac{EG_{,v}-F(2F_{,v}-G_{,u})}{2(EG-F^2)}.
\end{gather}  
\end{subequations}
To express~\eqref{eq:nA} in terms of the physical components we use the following identities 
\be
A_j=g_{ij} A^i=\sum_i \frac{g_{ij}}{h_{(i)}} A_{(i)},\qquad {\cal G}_{(ij)}=h_{(i)}h_{(j)} g^{ik} g^{jl} {\cal G}_{kl}\,\hat\bi_{(j)}\otimes\hat\bi_{(i)},
\ee
yielding
\begin{multline}\label{eq:boxdA}
\boxed{(\nabla \bA)_{(ij)}} = h_{(i)}h_{(j)}g^{ik}g^{jl}\bigg(\p_k A_l -\sum_m \Gamma_{kl}^m A_m\bigg) =\\=\boxed{h_{(i)}h_{(j)}g^{ik}g^{jl}\bigg(\p_k \sum_s\frac{g_{sl}}{h_{(s)}} A_{(s)} - \sum_m\Gamma_{kl}^m \sum_s \frac{g_{sm}}{h_{(s)}} A_{(s)}\bigg)\,\hat\bi_{(j)}\otimes\hat\bi_{(i)}}.
\end{multline}
Finally we compute the covariant derivative of $\bi_1$ and $\bi_2$ in the basis $\hat\bi_{(i)}$ with the help of {\it Mathematica}  and using
\be
\bi_1=\underbrace{1}_{A_{(1)}}\hat\bi_{(1)}, \qquad \bi_2= \underbrace{-\frac F{\sqrt{EG-F^2}}}_{A_{(1)}}\, \hat\bi_{(1)} +\underbrace{\sqrt{\frac{EG}{EG-F^2}}}_{A_{(2)}}\,\hat\bi_{(2)}. 
\ee 
The expressions are too bulky to reproduce here. The results~\eqref{eq:ne1} and \eqref{eq:ne2} are recovered for $F\equiv0$. More importantly, below we list the components of the tensor $\nabla\bn$~\eqref{eq:a14} within the considered general parametrization 
\begin{subequations}
\begin{align}
(\nabla\bn)_{11}=a_1&=\frac{\sin\alpha \big[E \big(E_{,v}-2 (F_{,u}+ \alpha_{,u}\sqrt{EG-F^2})\big)+F E_{,u}\big]}{2 E^{3/2} \sqrt{EG-F^2}},\\
(\nabla\bn)_{12}=a_2&=-\frac{\sin\alpha \big[E^2 \big(2 \alpha_{,v}\sqrt{EG-F^2} +G_{,u}\big)-2 F E \big(F_{,u}+\alpha_{,u}\sqrt{EG-F^2} \big)+F^2 E_{,u}\big]}{2 E^{3/2} (EG-F^2)},\\
(\nabla\bn)_{21}=a_3&=\frac{1}{2} \cos\alpha \bigg(\frac{2 \alpha_{,u}}{\sqrt E}-\frac{E (E_{,v}-2 F_{,u})+F E_{,u}}{\sqrt{E^3(EG-F^2)}}\bigg),\\
(\nabla\bn)_{22}=a_4&=\frac{1}{2} \cos\alpha \bigg(\frac{2 (E \alpha_{,v} -F \alpha_{,u})}{\sqrt{E (EG-F^2)}}+\frac{E^2 G_{,u}-2 F E F_{,u}+F^2 E_{,u}}{E^{3/2} (EG-F^2)}\bigg).
\end{align}
\end{subequations}
Replacing $a_i$ and~\eqref{eq:EFG} into~\eqref{eq:ftpm} gives the tetratic free energy for superspheroid.


\end{document}